\documentclass[5p]{elsarticle}

\usepackage{float}
\usepackage{hyperref}
\usepackage{amsmath}
\usepackage{amssymb}

\begin{document}

\begin{frontmatter}

\title{Design and Operational Experience of a Microwave Cavity Axion Detector for the $20-100\: \mu$eV Range}

\author[berk]{S. Al Kenany}
\author[col]{M.A. Anil}
\author[berk]{K.M. Backes}
\author[yale]{B.M. Brubaker}
\author[yale]{S.B. Cahn}
\author[llnl]{G. Carosi}
\author[yale]{Y.V. Gurevich}
\author[col]{W.F. Kindel}
\author[yale]{S.K. Lamoreaux}
\author[col]{K.W. Lehnert}
\author[berk]{S.M. Lewis}
\author[col]{M. Malnou}
\author[col]{D.A. Palken}
\author[berk]{N.M. Rapidis}
\author[berk]{J.R. Root}
\author[berk]{\corref{cor1}M. Simanovskaia}
\ead{simanovskaia@berkeley.edu}
\author[berk]{T.M. Shokair}
\author[berk]{I. Urdinaran}
\author[berk]{K.A. van Bibber}
\author[yale]{L. Zhong}

\address[berk]{Department of Nuclear Engineering, University of California Berkeley, Berkeley CA, 94720 USA}
\address[col]{JILA and the Department of Physics, University of Colorado and National Institute of Standards and Technology, Boulder CO, 80309 USA}
\address[yale]{Department of Physics, Yale University, New Haven CT, 06511 USA}
\address[llnl]{Physical and Life Sciences Directorate, Lawrence Livermore National Laboratory, Livermore CA, 94551 USA}

\cortext[cor1]{Corresponding author}

\begin{abstract}
We describe a dark matter axion detector designed, constructed, and operated both as an innovation platform for new cavity and amplifier technologies and as a data pathfinder in the $5 - 25$ GHz range ($\sim20-100\: \mu$eV). The platform is small but flexible to facilitate the development of new microwave cavity and amplifier concepts in an operational environment. The experiment has recently completed its first data production; it is the first microwave cavity axion search to deploy a Josephson parametric amplifier and a dilution refrigerator to achieve near-quantum limited performance.
\end{abstract}

\begin{keyword}
\texttt{axion, dark matter, Josephson Parametric Amplifier, microwave cavity, standard quantum limit, superconducting magnet}
\end{keyword}

\end{frontmatter}


\section{Introduction}

The axion is a hypothetical pseudoscalar arising from the Peccei-Quinn mechanism to protect the strong interaction from CP-violating effects. Also, an axion in the $1-100\; \mu$eV mass range is a compelling dark matter candidate. A comprehensive review of the particle physics of the axion, its cosmological and astrophysical significance, and experimental searches for it can be found in Ref.~\cite{Gra15}.

\begin{figure}[h]
\centerline{\includegraphics[width=.5\textwidth]{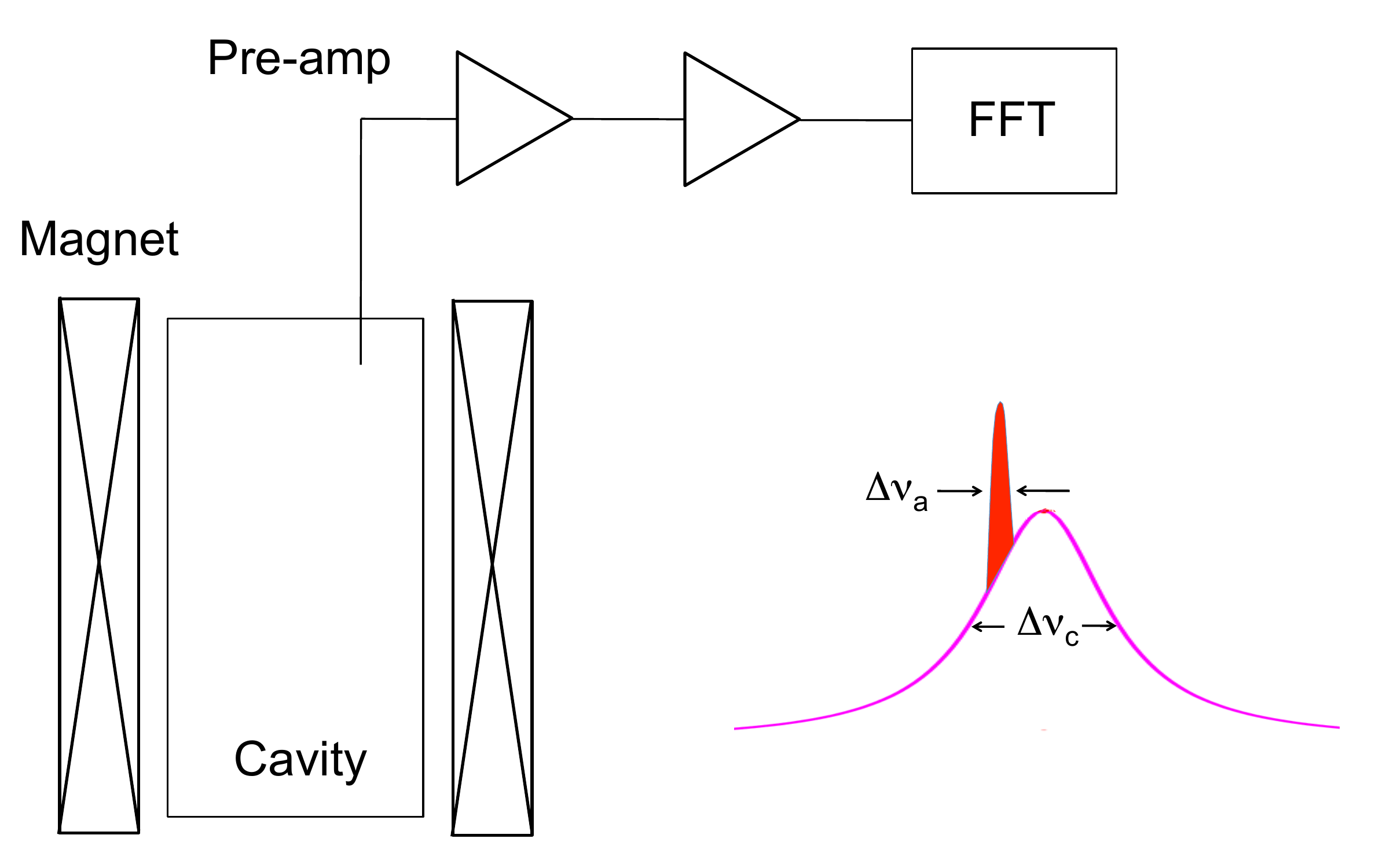}}
\caption{Schematic of the microwave cavity search for dark matter axions. The axion signal is designated by the narrow peak (red) within the bandpass of the cavity (pink).}\label{Fig1}
\label{fig:schematic}
\end{figure}

The axion, like the $\pi^0$, can couple to two photons, one of which may be virtual. Sikivie thus proposed a practical detection strategy based on the resonant conversion of the dark matter axion to a single microwave photon carrying its full energy (mass + kinetic) in a high-Q cavity permeated by a strong magnetic field~\cite{Sik83,Sik85,kra1985}. The resonant conversion condition is that the axion mass $m_a$ is close to the resonant frequency $\nu_c$ of a cavity mode with an appropriate spatial profile; more precisely, $\left|\delta\nu\right| = \left|\nu_c-m_ac^2/h\right| \lesssim \Delta\nu_c$, where $\Delta\nu_c$ is the linewidth of the mode. The axion kinetic energy distribution is generally assumed to be Maxwellian, parametrized by the virial velocity $\left<v^2\right>^{1/2}\sim10^{-3}\,c$. The signal is thus monochromatic to a part in $10^6$, with linewidth $\Delta\nu_a = m_a\left<v^2\right>/h \ll \Delta\nu_c$; see Fig.~\ref{fig:schematic}.

The axion conversion power in such a microwave cavity detector is

\begin{equation}
\begin{split}
P_{\text{sig}} = & \left(g_\gamma^2\frac{\alpha^2}{\pi^2} \frac{\hbar^3c^3\,\rho_a}{\Lambda^4}\right) \times \\ 
	& \left(\frac{\beta}{1+\beta}\omega_c\frac{1}{\mu_0}B_0^2VC_{mn\ell}Q_L\frac{1}{1+\left(2\delta\nu/\Delta\nu_c\right)^2}\right).\\
\end{split}
\label{eq:power}
\end{equation}

Eq.~\eqref{eq:power} is valid in any self-consistent set of units; the two sets of parentheses contain theory and detector parameters, respectively. On the theoretical side, $\alpha$ is the fine-structure constant, $\Lambda=77.6$~MeV encodes the dependence of the axion mass on hadronic physics, and the local axion dark matter density $\rho_a$ and coupling constant $g_\gamma$ are the parameters that experiment can constrain.\footnote{The value of $\Lambda$ used in our analysis was obtained from chiral perturbation theory~\cite{Sik85}. Note also that $\Lambda^4 = \chi(T=0)$, where $\chi$ is the QCD topological susceptibility which may be calculated on the lattice. The most recent lattice calculation, in Ref.~\cite{bor2016}, obtained $\Lambda=75.6~\text{MeV}$, which would result in an 11\% enhancement of the signal power.} It is conventional to fix $\rho_a=0.45$~GeV/cm$^{3}~$\cite{Pen00} and quote limits on $g_\gamma$, a model-dependent dimensionless coupling that is related to the physical coupling $g_{a\gamma\gamma}$ appearing in the axion-photon Lagrangian by $g_{a\gamma\gamma}=\left(g_\gamma\alpha/\pi\Lambda^2\right)m_a$. In the KSVZ (DFSZ) benchmark axion model, $g_\gamma=-0.97$ ($0.36$), independent of the axion mass.

The experimental parameters in Eq.~\eqref{eq:power} include the magnetic field strength $B_0$, the cavity volume $V$, and several factors characterizing the cavity mode, with $\omega_c=2\pi \nu_c$. The coupling between the cavity mode and the receiver used to detect the signal, parametrized by $\beta$, reduces the quality factor from $Q_0$ to $Q_L=Q_0/\left(1+\beta\right)$. The form factor $C_{mn\ell}$ parametrizes the overlap between the cavity mode and the external magnetic field. For the cylindrical geometry commonly used in cavity axion detectors, $C_{mn\ell}$ may be written

\begin{equation}
C_{mn\ell} = \frac{\left(\int\mathrm{d}^3\mathbf{x}\,\mathbf{\hat{z}}\cdot\mathbf{e}^*_{mn\ell}(\mathbf{x})\right)^2}{V\int\mathrm{d}^3\mathbf{x}\, \epsilon\left(\mathbf{x}\right) \, \left|{\mathbf{e}_{mn\ell}(\mathbf{x})}\right|^2},
\label{eq:c010}
\end{equation}

\noindent where $\mathbf{e}_{mn\ell}(\mathbf{x})$ is the normalized electric field profile of the mode, $\epsilon(\mathbf{x}) = 1$ is the dielectric constant inside the cavity, and $B_0$ is axial and homogeneous. Nodes in the electric field profile lead to cancellations in the form factor, and thus it is only appreciable for low-order TM$_{0n0}$ modes.

Inserting typical values for the detector described in this work, which are close to the limits of present technology, the peak signal power is $P_{\text{sig}}\sim5\times10^{-24}$~W at KSVZ coupling, a factor of $10^9$ below room-temperature thermal noise power in a typical cavity bandwidth. Thus cryogenic operation and a low-noise receiver are necessary for any practical realization of a cavity axion detector. The intrinsic spectral resolution of the linear receivers used in cavity axion searches to date implies that the relevant noise bandwidth is $\Delta\nu_a$ rather than $\Delta\nu_c$, and moreover that a measurement at any given cavity frequency simultaneously probes $\sim\Delta\nu_c/\Delta\nu_a$ independent values of the axion mass. Further improvement in the signal-to-noise ratio (SNR) may then be obtained by averaging the cavity noise for a time $\tau$. The preceding discussion is formalized in the Dicke radiometer equation \cite{dicke1946}, in which the SNR is defined as

\begin{equation}
\Sigma = \frac{P_{\text{sig}}}{k_BT_{\text{sys}}}\sqrt{\frac{\tau}{\Delta\nu_a}}.
\label{eq:SNR}
\end{equation}

For any phase-insensitive linear receiver the system noise temperature $T_{\text{sys}}$ may be written

\begin{equation}\label{eq:noise}
k_BT_{\text{sys}} = h\nu N_{\text{sys}} = h\nu\left(\frac{1}{e^{h\nu/k_BT} -1} + \frac{1}{2} + N_A\right),
\end{equation}

\noindent where the three additive contributions correspond respectively to a blackbody gas in equilibrium with the cavity at temperature $T$, the zero-point fluctuations of the blackbody gas, and the input-referred added noise of the receiver. The price we pay for spectral resolution is the quantum limit $N_A \geq 1/2$ on the added noise of any phase-insensitive linear receiver~\cite{caves1982}, which together with the second term in Eq.~\eqref{eq:noise} implies a ``standard quantum limit'' $N_{\text{sys}} \geq 1$ for microwave cavity axion detection. These limits imply that units of quanta, used throughout this work, are more appropriate than temperature units for sufficiently low-noise receivers.

The axion mass is a priori unknown, so the cavity must be tuned in small discrete steps after each measurement interval $\tau$; the SNR for any given value of the axion mass is then given by a quadrature sum of terms with the form of Eq.~\eqref{eq:SNR}. An average scan rate is obtained by solving for $1/\tau$ and multiplying by the frequency step size. With the simplifying assumption that the detector parameters remain approximately constant over its tuning range, the scan rate is

\begin{equation}
\begin{split}
\frac{\mathrm{d}\nu}{\mathrm{d}t} \approx \frac{4}{5}\frac{Q_LQ_a}{\Sigma^2} & \left(g_\gamma^2\frac{\alpha^2}{\pi^2}\frac{\hbar^3c^3\rho_a}{\Lambda^4} \right)^2 \times \\
& \left(\frac{1}{\hbar\mu_0}\frac{\beta}{1+\beta}B_0^2VC_{mn\ell}\frac{1}{N_{\text{sys}}}\right)^2,\\
\end{split}
\label{eq:scan}
\end{equation}

\noindent where $Q_a=\left(\left<v^2\right>/c^2\right)^{-1}$ is the ``quality factor'' of the axion signal, and the factor of 4/5 comes from a sum over the squared Lorentzian factors characterizing the effect of the axion's changing detuning from the cavity mode. This is a good approximation for any frequency step size $\lesssim\Delta\nu_c/2$; qualitatively, a smaller step size implies that the same sensitivity can be achieved with smaller $\tau$, and the two effects cancel out in the scan rate. Note also that, while Eq.~\eqref{eq:power} is maximized at critical coupling ($\beta=1$) for a given $Q_0$, the scan rate is maximized for an overcoupled cavity with $\beta=2$. Eq.~\eqref{eq:scan} is the most useful figure of merit for the cavity axion search.

The detector described in this work builds upon experience from the Axion Dark Matter eXperiment (ADMX), a larger platform designed to operate initially in the sub-GHz range. ADMX entered into operation in 1996 with a physical temperature of $T \approx$ 1.5~K~\cite{Pen00}. The readout chain, initially led by a high electron mobility transistor amplifier (HEMT)~\cite{Pen00}, was later upgraded to include a microstrip-coupled SQUID amplifier (MSA)~\cite{Asz11}. A dilution refrigerator has recently been incorporated into the setup, and commissioning is underway. 

Our detector was conceived as both an innovation platform and a data pathfinder for the microwave cavity axion search in the $5-25$ GHz range ($\sim20-100\; \mu$eV axion masses; 1 GHz = 4.136 $\mu$eV). The next decade in frequency brings new challenges for both microwave cavities and receivers. On the cavity side, preserving the aspect ratio, the volume of the cavity $V \sim \nu^{-3}$, resulting in an increasing penalty in conversion power with increasing frequency. This requires insightful cavity designs to maintain useful volume at higher frequencies without sacrificing the form factor or incurring an unacceptable density of intruder TE modes in the spectrum. The absolute machining and alignment tolerances will also become tighter for smaller structures, to avoid mode localization. Another challenge is that MSAs do not work well above a gigahertz; to circumvent this problem we have introduced Josephson parametric amplifiers (JPA) that are ideal for the 5~GHz range. Together with a dilution refrigerator (DR), built into our design from the beginning, the JPA allows us to push down towards the Standard Quantum Limit and thus remain competitive with lower-frequency limits despite the loss of sensitive volume. 

At the same time, this frequency range presents an opportunity to deploy entirely novel technologies that promise dramatic new capability for the microwave cavity search. A technology that will be investigated in the near future is a receiver based on squeezed states of the vacuum to circumvent the Standard Quantum Limit. Innovations to be tested in the microwave cavity domain include photonic band gap resonators, designed to eliminate the spectrum of TE modes which otherwise mix with the TM$_{010}$ mode of interest as it is tuned, resulting in a loss of sensitivity at mode crossings and making it more difficult to track the TM$_{010}$ mode. Other schemes to be investigated are the application of distributed Bragg reflector based resonator schemes, and thin-film Type II superconducting coatings to improve the quality factor $Q$ of the cavity.

\section{Description of Experimental Setup}

\begin{figure*}[h]
\centerline{\includegraphics[width=1\textwidth]{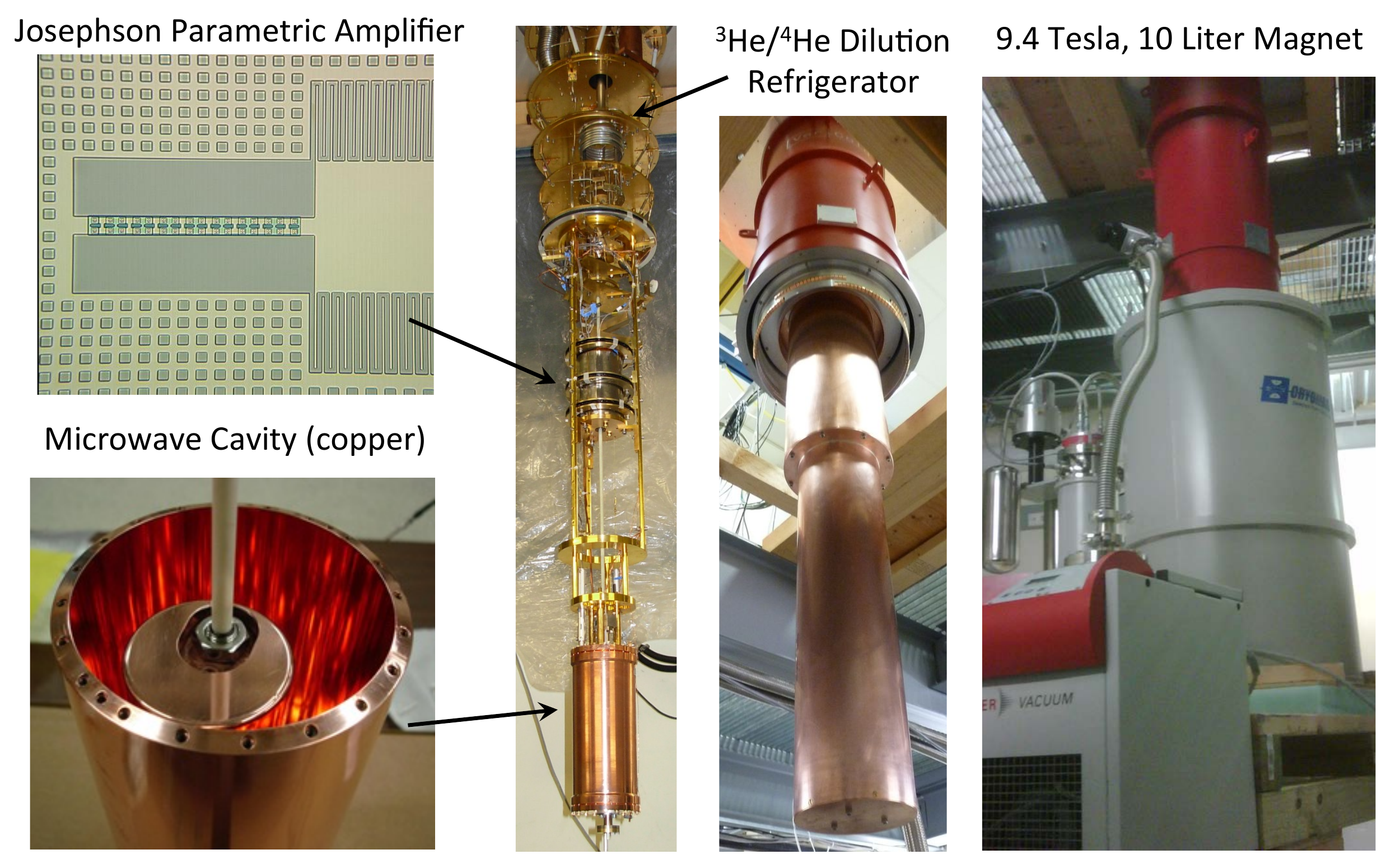}}
\caption{Overview of the experiment and its integration.}\label{Fig2}
\label{fig:overview}
\end{figure*}

Fig.~\ref{Fig2} presents an overview of the experiment, sited at the Wright Laboratory of Yale University, and its integration. The microwave cavity and the magnetically-shielded canister housing the JPA are assembled on a gantry suspended from the dilution refrigerator. The gantry assembly is lowered by crane into the bore of the magnet, which is located in a room below the floor level of the lab containing the electronics and computer control. 

The DR precooling system and the magnet cryogenic system use pulse tube and Gifford-McMahon cryocoolers, respectively. This allows operation of our detector without external liquid cryogens, providing major simplification of operations, and a substantial reduction in operating costs. However, such a cryogen-free system relies on an uninterrupted supply of electrical power as the consequences of a magnet quench can be severe (see section~\ref{sec_magnet}).  

The individual systems are described in detail below.

\subsection{Dilution Refrigerator}

The DR was manufactured in 2007 by VeriCold Technologies (subsequently acquired by Oxford Instruments), and was among the first commercially available external-cryogen-free DRs. The only major modification was the replacement of the VeriCold pulse tube with a Sumitomo RDK-415D Gifford-McMahon cryocooler (in retrospect this was a poor choice because of excessive vibration). In addition, instead of using the DR's Lakeshore temperature bridge to control the mixing chamber plate, we have incorporated a custom temperature PID-feedback controller at the mixing chamber level, resulting in reduced delay and improved performance.

The cooling power of the DR is about 150 $\mu$W at 127 mK, which is the system operating temperature. This operating temperature was chosen because we observed strong modulation of the JPA gain by vibrational fluctuations at lower temperatures; the temperature-dependence appears to originate in the circulator's magnetic shielding, whose susceptibility and heat capacity both fall below 4 K. In addition, operating at higher temperature results in reduced temperature excursions from mechanical motions and actuation of the microwave switch, due to the increased heat capacity of the mixing chamber plate. The cavity, JPA, mixing chamber, and still temperatures are monitored with calibrated ruthenium oxide sensors.  

The support gantry is a tripod with copper alloy legs and copper rings at both ends, which are clamped to the bottom of the DR's mixing chamber plate and to the upper endcap of the cavity, respectively. The equilibration time for changes in the mixing chamber temperature to propagate to the cavity temperature (measured at the lower endcap) is on the order of a few minutes. This might be surprising as the cavity is constructed of stainless steel, which has very low thermal conductivity. However, the cavity has a thick Cu plating ($0.125$ mm) which provides a sufficient thermal link for a short equilibration time. 

The cavity and gantry are thermally shielded with an extension of the DR still shield. This extension comprises two demountable sections with dimensions 17.8 cm OD $\times \, 45$ cm and 13.0 cm OD $\times \, 53.3$ cm for the upper and lower sections, respectively. At the bottom of the still shield extension, there is a G-10 fiberglass disk attached via a stainless steel rod, which centers the shield extension and maintains the $0.5$ cm gap between the still shield and magnet bore. The still shield extension was originally constructed from $0.16$ cm sheet copper with welded seams, but that was deformed when we experienced a magnet quench (more information in section~\ref{sec_magnet}). Subsequently, we constructed a replacement using heavily plated stainless steel ($0.125 - 0.250$ mm), prompted by our experience that a heavy copper plating provides a good thermal link.

The DR and magnet share a single common insulating vacuum space. Extensions of the 4 K and 77 K DR stages overlap the corresponding stages at the top of the magnet, and thermal contact is provided by finger stock that presses between the extensions and the corresponding magnet thermal stages and also by blackbody radiation between the overlapping surfaces. This design allows the DR/cavity/still shield assembly to be easily inserted into and removed from the magnet by use of an overhead hoist. The gas flow and electronic control lines are long enough that only a few cables must be disconnected to insert or remove the cryostat. The DR itself is supported from above, and the magnet is on a cart equipped with jack screws. When the DR is lifted, the magnet can be rolled away and the DR lowered again to its support, allowing easy and safe access to the assembly below the DR. To insert the DR, the magnet is rolled to the appropriate location, the DR and extension are lowered into the magnet, and the jack screws are used to raise the magnet until the corresponding DR and magnet cryostat vacuum flanges contact, after which the flanges are bolted together.

The DR operating alone has a cooldown time of about 14 hours; with the gantry, cavity, and magnet, the cooldown time to 127 mK cavity temperature is 72 hours. 

\subsection{Magnet}
\label{sec_magnet}

The magnet and its controller were designed and manufactured by Cryomagnetics, Inc., based on an existing design for a gyrotron system with similar field requirements. As mentioned already, the magnet is a pulse-tube cooled system, requiring no external cryogens or helium gas, and operates in a persistent mode. The magnetic field is ramped up to a maximum of 9 T over 8~hours; the rate is limited by heating of the magnet itself and by eddy current heating of the cavity/gantry/still shield assembly. An interesting feature of the heating due to the ramping is that the heat load inferred from the rise in cavity temperature is largest when the field is between about 4 and 7 T, which we attribute to electrons being polarized in the stainless steel cavity body.
  
Our design specification was that the field needs to be homogenous at a level such that the perpendicular (radial) field at the inner surface of the cylindrical cavity barrel is less than 50 G. For future versions of this axion detector, we are interested in coating this inner surface with superconducting thin films to improve $Q_0$. The mentioned design specification is a requirement for a thin film superconducting layer to remain effective in the presence of a strong magnetic field. The idea is that the vortices formed by the perpendicular component of the field would be of sufficiently low density and mobility to not contribute significantly to radiofrequency power loss in the film \cite{tanner}. The modification of the gyrotron design includes a set of superconducting coils in series with the main magnet that cancel the axial field over a 15.2 cm region to less than 50~G. This is the location of the JPA and its shield.

The magnet bore diameter is 14 cm inside the main solenoid and steps up to 18.4 cm further up, where the bucking coils are supported. The effective and homogeneous region in the field maximum extends axially over at least 25 cm.

An unfortunate disadvantage of a cryogen free system is that it relies on an uninterrupted source of power. The time to quench after a power loss is about 4 minutes for our system. In principle, Yale has fast acting emergency backup power, but due to system upgrades and other issues, this emergency power was not available in early March 2016, when an unscheduled power outage resulted in a magnet quench, causing significant mechanical/structural damage to the DR and coaxial microwave lines. The repairs of this damage took two months, and fortunately none of the DR's $^3$He was lost; in spite of the mechanical damage, the gas flow lines were in no way compromised. Most of the forces due to this quench were caused by the large copper rings at the lower part of the gantry, and by the copper still shield extension. As noted above, we have replaced the copper still shield extension with another made from copper-plated stainless steel. In future designs, most of the gantry components will be constructed from copper-plated stainless steel, which should lead to more than a 100-fold reduction in forces during a quench.

\begin{figure}[h]
\centerline{\includegraphics[width=.5\textwidth]{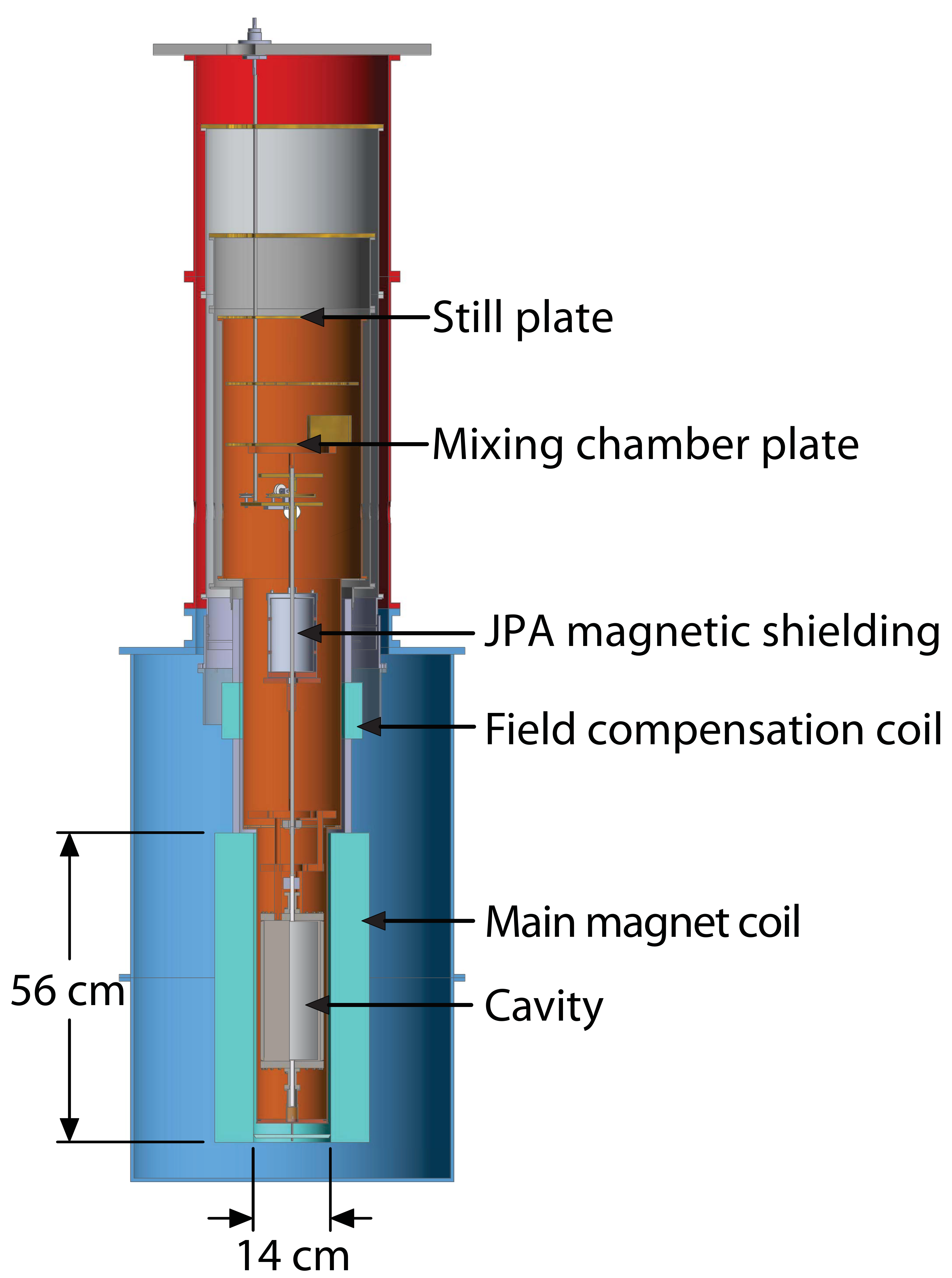}}
\caption{Layout of the experiment. The red volume is the DR vacuum shield, the blue volume is the magnet body, and the orange shaded region is the volume bounded by the still shield and its extensions. The magnet's 70 K and 4 K shields are not shown below their interface with the fridge shields.}\label{Fig3}
\label{fig:magnetschem}
\end{figure}

\subsection{Microwave cavity\label{sub:cavity}}

The initial microwave cavity for the experiment consists of a right circular cylinder of stainless steel, electroplated with oxygen-free high conductivity (OFHC) copper.  After annealing, the cavity achieves a near-theoretical maximum value of the quality factor $Q$, as limited by the anomalous skin depth~\cite{Kit05}, 

\begin{equation}
\delta_{\text{anom}} = \left( \frac{\sqrt{3}\, c^2\, m_e\, v_F}{8\pi^2\, \omega\, n\, e^2} \right)^{1/3},
\end{equation}

\noindent where $m_e$ is the mass of the electron, $n$ the electron density in the metal, $v_F$ the Fermi velocity, and $\omega$ the angular frequency. For copper, $n = 8.5\times 10^{22}$~cm$^{-3}$ , and  $v_F  = 1.57\times 10^{8}$~m/s.   

\begin{figure}[t]
\centerline{\includegraphics[width=.5\textwidth]{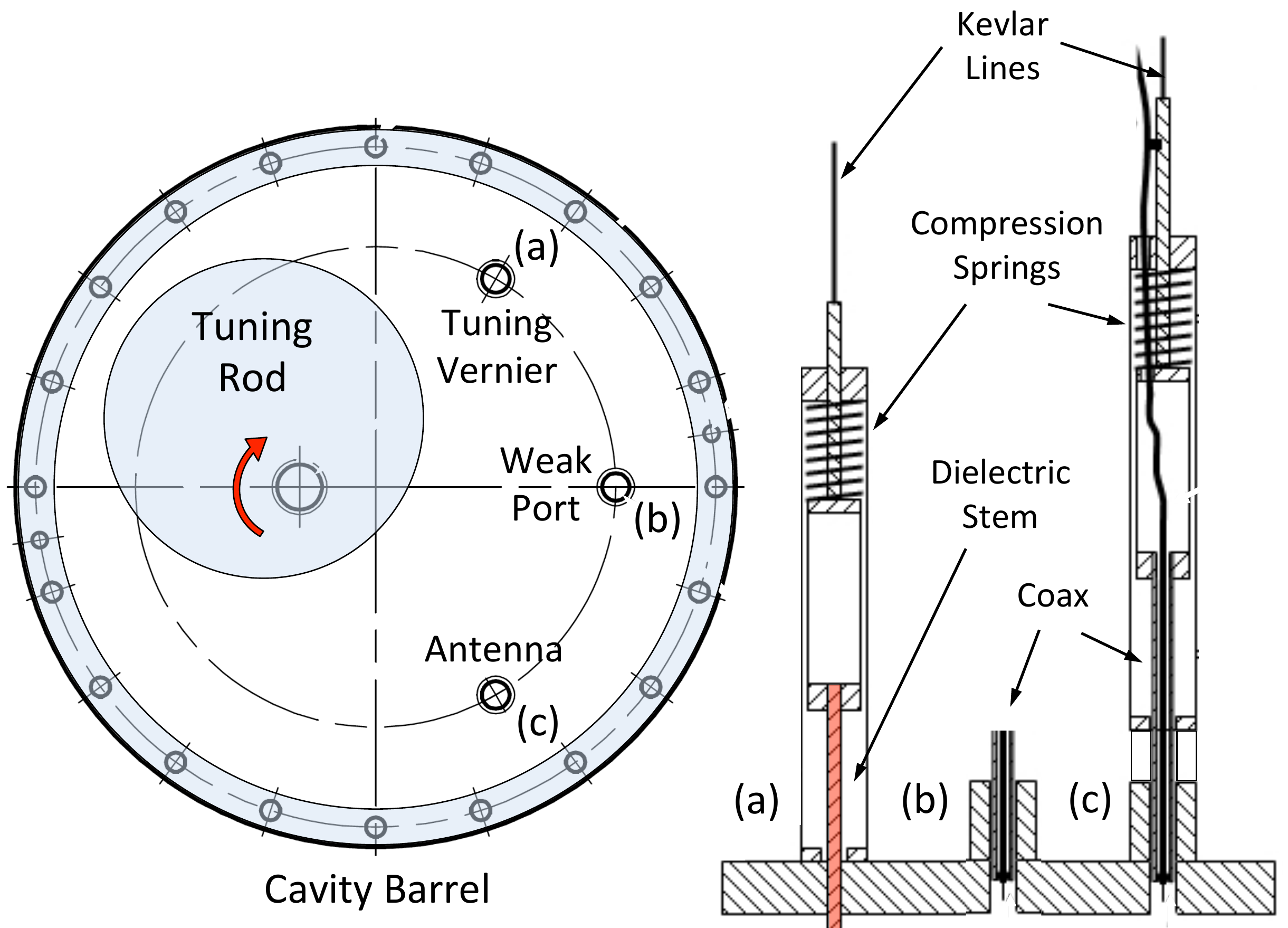}}
\caption{Top and side view of the cavity upper endcap, detailing the positions and design of the main tuning rod, tuning vernier, signal injection port, and the variable-coupling antenna.}
\label{fig:vernier}
\end{figure}

The cavity employed for the initial data run was 25.4~cm in height and 10.2~cm in diameter. A single large diameter rod, $\O$ 5.1~cm, was used to tune the cavity (see Fig.~\ref{Fig2}). The rod pivoted around an off-axis ceramic axle, such that its radial position could be adjusted from touching the wall to centered within the cavity. This produced a dynamic range of the TM$_{010}$-like mode of $3.6 - 5.8$~GHz. To make finer frequency steps, we used a tuning vernier, a 3.2 mm diameter alumina rod, of variable insertion depth into the cavity. With the tuning rod centered, the cavity is more properly described as an annular cavity, which has the largest volume for a TM$_{010}$-like mode at a given frequency. The upper endcap had ports for two antennas, one with weak coupling used to measure the cavity's response in transmission and one whose insertion depth (and thus coupling $\beta$) was variable. The penetrations and services of the cavity are shown in Fig.~\ref{fig:vernier}. Around 5.8 GHz, typical values were $C_{010} \sim 0.5$ for the form factor and $Q_0 \sim 30,000$ for the unloaded cavity Q.

\begin{figure}[t]
\centerline{\includegraphics[width=.45\textwidth]{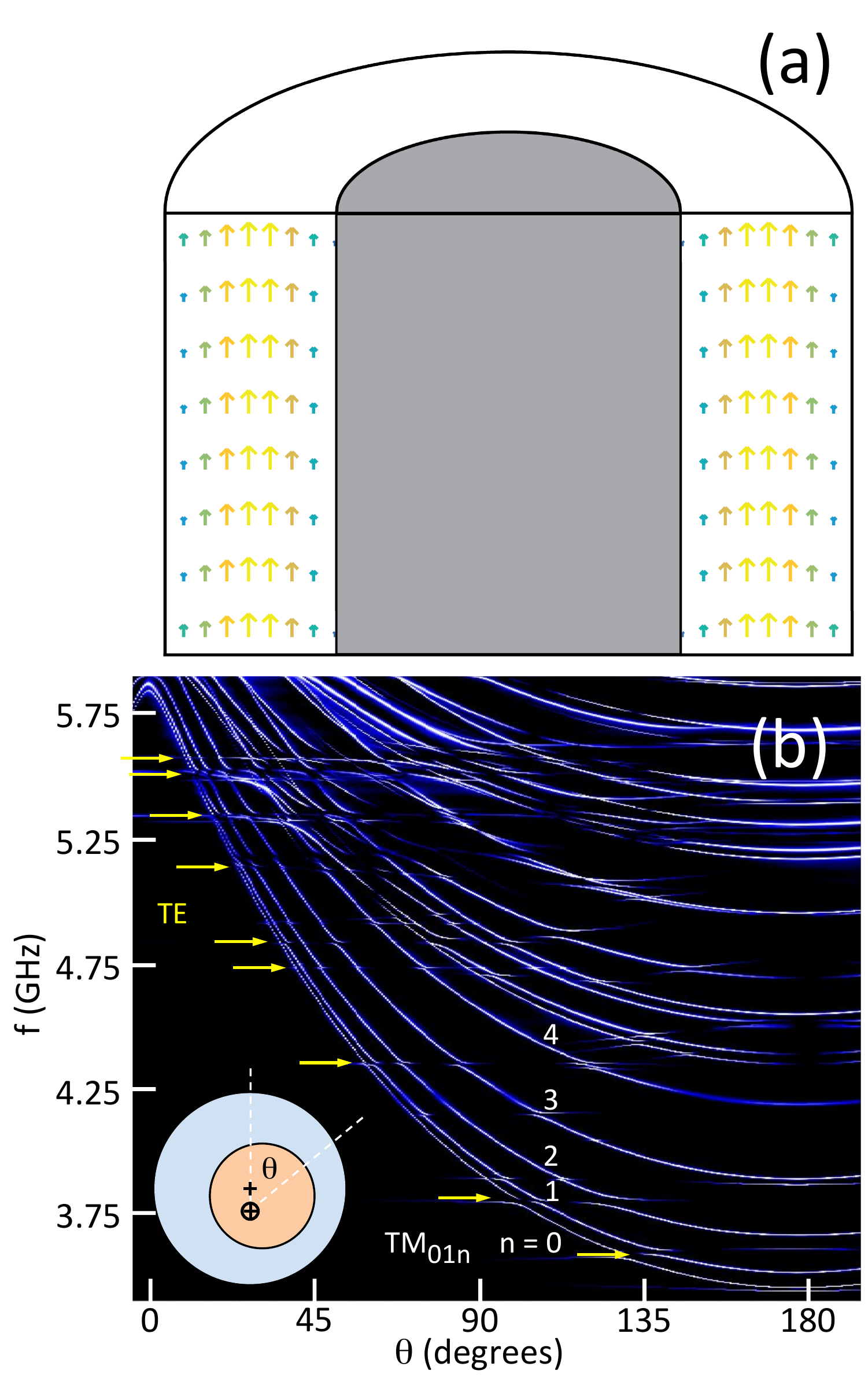}}
\caption{(a) CST Microwave Studio simulation of the TM$_{010}$-like mode with the tuning rod in the middle of the cavity. (b) Measured mode map of the cavity, as a function of the pivot angle of the tuning rod within the cavity. The frequencies of the TM$_{01n}$ modes decrease steeply with increasing radial distance of the tuning rod from cavity center. TE modes (indicated by yellow arrows) become apparent approaching mode crossings with TM modes; the TE mode frequencies are largely insensitive to the position of the tuning rod.}
\label{fig:CST}
\end{figure}

\begin{figure}[h]
\centerline{\includegraphics[width=.5\textwidth]{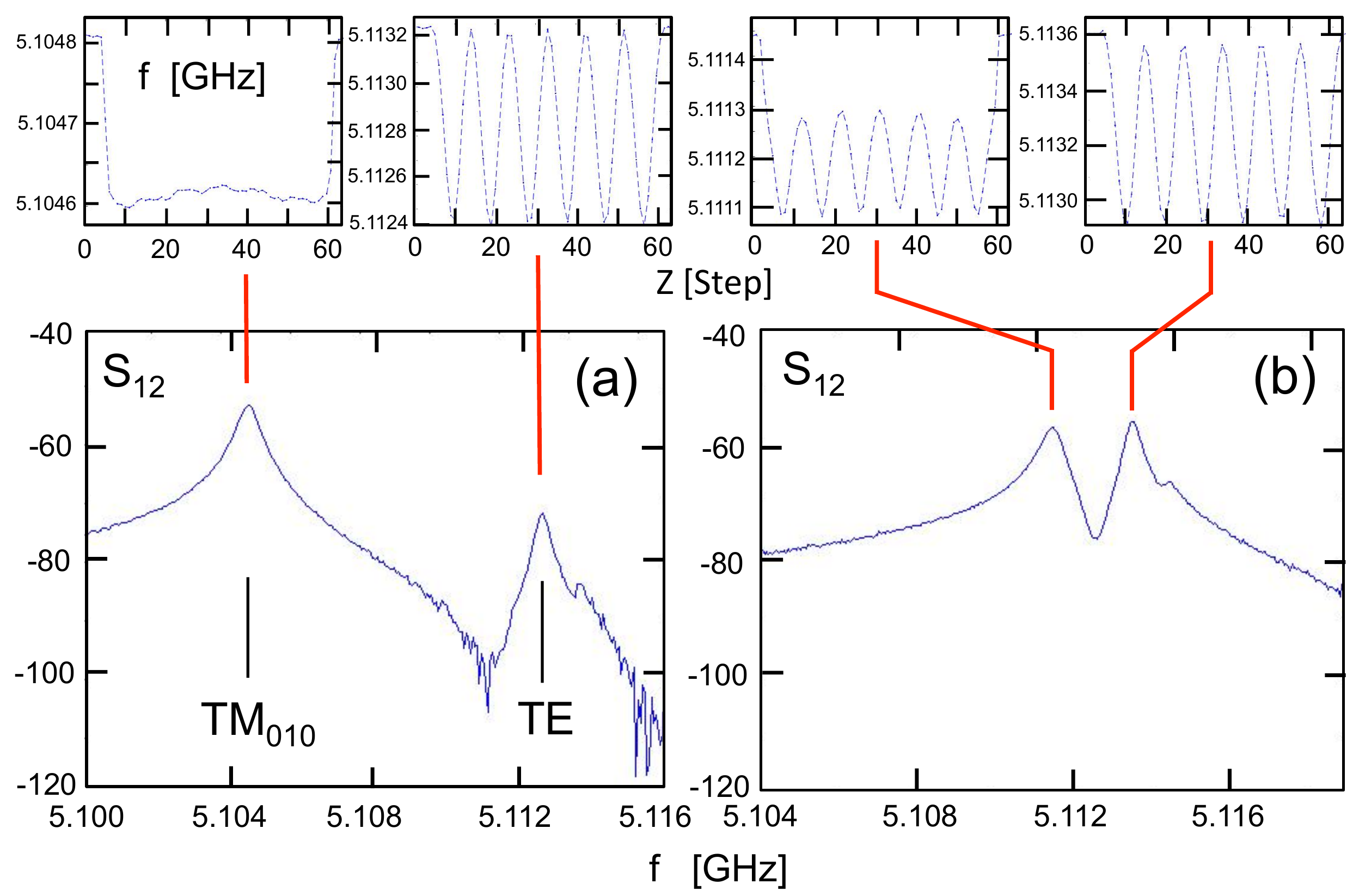}}
\caption{Cavity spectra for two different rod positions, as the TM$_{010}$-like mode approaches a crossing with a high-order TE mode. For each rod position, the plots above the main panels show the shifts in frequency of the two modes, as a small dielectric bead is translated axially through the cavity. The frequency shift at each axial position Z is proportional to the square of the local electric field according to perturbation theory (see Eq.~\ref{beadpert}), thus elucidating the nature of the mode. (The spectra in the main panels are with the bead removed from the cavity.) (a)  At a separation of 8 MHz, the purity of the TM$_{010}$-like mode is confirmed by the flat axial field profile shown on the top left. (b) At a separation of 2 MHz, the lower-frequency mode now clearly reflects a strong admixture of both a high-order TE mode along with the TM$_{010}$-like mode.}
\label{fig:modecross}
\end{figure}

Our cavity development program has involved precision metrology and modeling in simulation environments such as CST Microwave Studio. Fig.~\ref{fig:CST} displays a simulation of the TM$_{010}$-like mode and a mode map of the cavity. Another innovation of this axion search is the use of the bead perturbation method to measure field profiles in a microwave cavity. To date, these have been longitudinal profiles along the full length of the cavity at a single transverse position, derived by translating a small dielectric bead (alumina, $\epsilon \sim 10$) on a Kevlar line controlled by a stepper motor and some pulleys. For a dielectric bead, the shift in mode frequency in perturbation theory is given by:

\begin{equation}
\frac{\Delta \omega}{\omega} = \frac{-\left(\epsilon-1 \right)}{2} \frac{V_\text{bead}}{V_\text{cav}} \frac{E(r)^2}{\left< E(r)^2\right>_{\text{cav}}}
\label{beadpert}
\end{equation}

\noindent where $V_\text{bead}$ and $V_\text{cav}$ indicate the volume of the dielectric bead and cavity, respectively~\cite{Sla46}. For the TM$_{010}$-like mode, the electric field should be longitudinally invariant, and thus the shift in frequency as a function of longitudinal distance should be constant. In the microwave cavity axion search, the bead perturbation technique is useful in the design and testing of new resonators, to determine the symmetry of a mode, to confirm or correct alignment and machining tolerances to eliminate mode localization, and to identify mode-crossings regions where the modes have become substantially admixed, and thus the form factor C$_{mnl}$ of the TM$_{010}$-like mode has begun to diminish (Fig.~\ref{fig:modecross}). A detailed study of microwave cavities for the dark matter axion experiment bringing together metrology, simulation, and EM characterization will be the subject of a future publication. For the present, a few observations in relation to experiments an order of magnitude lower in frequency~\cite{Pen00, Asz11} may be useful.

First, for cavities of characteristic dimension $O$(10~cm), machining and alignment tolerances to avoid significant mode localization~\cite{Hag90} are approaching the limit of good machine shop standards and readily available components. Even the transverse play of the inner race of the bearing for the axle of the tuning rod, of order $50-75 \, \mu$m, can cause a tilt in the TM$_{010}$ axial electric field profile of several percent. Second, modeling revealed that for the large tuning rod used in this experiment (r/R = 0.5), the form factor $C_{010}$ only approached the value associated with the idealized case (i.e. the rod electrically joined to the top and bottom endcaps) for gaps $G <250\, \mu$m. Third, it is becoming clear that engineered features necessary for practical cavities (e.g. penetrations for the antenna coupling, diagnostic ports, etc.) have a significant effect on the mode, and are likely to be more important at higher frequencies given the larger relative size of mechanical and electrical components penetrating the cavity.

\subsection{Mechanical Controls}

The experiment has three mechanical systems: the rotary cavity tuning, the adjustment of the vernier, and the adjustment of the antenna.  All mechanical controls are based on Applied Motion Products HT-23-595D NEMA 23, high-torque, double-shaft stepping motors, driven by model 5000-235 STR2 microstepping controllers. One side of each stepping motor double shaft is coupled to a ten turn potentiometer that encodes the net rotation angle. The vernier and antenna stepping motors are coupled to the experiment using 10:1 worm gear reductions, while the cavity tuning motor is coupled directly. The rotary motions are coupled into the vacuum using three K.J. Lesker FMH-25A Dynamic O-Ring Shaft Seal feedthroughs.  

Within the cryostat, the antenna and vernier are controlled by 0.36 mm Kevlar thread lines that are wound directly onto the rotary feedthrough shaft that extends into the vacuum. At each stage of the DR, the lines pass through 3.2 mm ID, 2.5 cm long tubes that are thermally linked to the DR stage and serve as radiation shields. The lines are routed to the cavity by use of nylon pulleys with ball bearings (McMaster-Carr 3434T13) that have been ultrasonically cleaned to remove all lubricants that would freeze at low temperatures. It should be noted that Kevlar expands on cooling, and the spring tensioning of the lines must absorb the length change lest the lines loosen and possibly fall out of the pulley grooves.

The antenna and vernier are mounted on fixtures that allow motion along the cavity axis only. A spring is compressed when the antenna or vernier is pulled out of the cavity by the Kevlar thread, ensuring smooth and reversible motion.

Rotary motion for the tuning rod system is delivered to the mixing chamber level by use of a cryogenic G-10 tube (6.4 mm diameter, 0.79 mm wall). The ends of the G-10 tube are fitted with glued-in brass extensions which allow the use of set screws to couple to the feedthrough at the top, and to the mechanics at the mixing chamber end.  At each DR stage, a 2.5 cm section of brass tubing (6.4 mm ID, 0.4 mm wall) is glued to the G-10, and a loose-fitting brass tube (1.9 cm long, 0.4 cm ID) is slipped over each 2.5 cm section and coupled to the stage using copper braid.  At the 4 K stage, the G-10 tube inside has a brass blackbody radiation block. 

Just below the mixing chamber, a pulley and torsion spring system is used to transfer rotary motion from the upper G-10 tube to a cryogenic G-10 tube (also 6.4 mm) along the DR and magnet axis, supported at the mixing chamber end by a ceramic bearing. A kevlar line connects the brass extension of the upper shaft to the 10.2 cm diameter pulley, and the torsion spring ensures that the line is always pulled taut. The lower G-10 shaft runs through the JPA magnetic shield and down to the cavity, where a 1.4:1 anti-backlash gear reduction provides the final radial displacement required to couple to the tuning rod axle.

These mechanical systems have generally provided the levels of control needed for the antenna and vernier; all three required appropriate selection of the microstepping resolution and had negligible heat loads at the operating temperature. The cavity bearings had some stiction due to alignment issues, and this resulted in sometimes erratic positioning, and a slow drift to final equilibrium after stepping. In operation of the experiment, the fine frequency control was done using the vernier, with less frequent larger rotations of the tuning rod which required up to 15 minutes of waiting for the rod to come to its equilibrium position. We plan to eliminate this problem by improving the cavity mechanics and replacing the room-temperature tuning rod drive system with an AttoCube piezoelectric rotator at base temperature.

\subsection{Josephson Parametric Amplifier\label{sub:jpa}}

\begin{figure}[h]
\centering\includegraphics[width=0.45\textwidth]{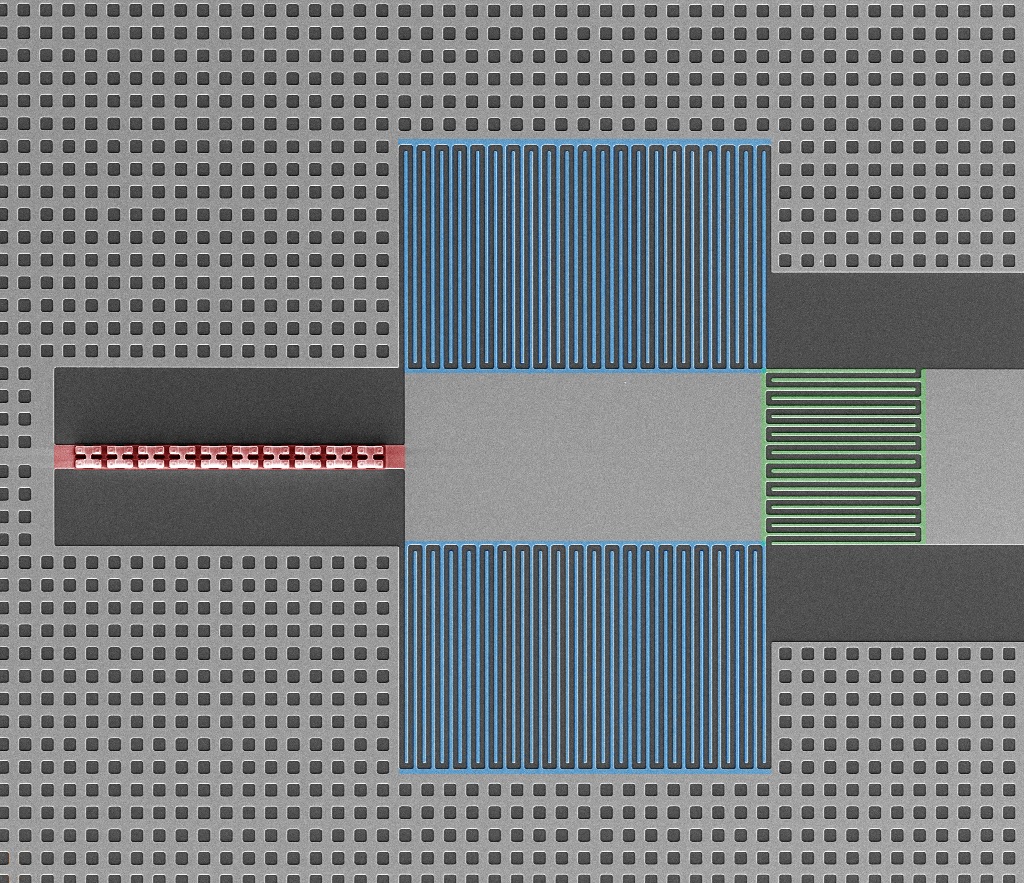}
\caption{Microphotograph of JPA circuit. The SQUID array (approximately 150 $\mu$m long) is highlighted in red on the left; the circuit's resonance is determined by the SQUID inductance and the geometric capacitance (blue). The circuit is coupled to a $50~\Omega$ transmission line through a smaller capacitance (green). The surrounding superconducting ground plane is waffled in order to pin magnetic flux vortices in place and keep them from the SQUID array.\label{fig:jpa_layout}}
\end{figure}

A JPA is a nonlinear LC resonator capacitively coupled to a transmission line with an array of SQUIDs playing the role of the nonlinear inductance (see Fig.~\ref{fig:jpa_layout}). By applying a DC flux through the SQUID loops, the resonant frequency can be tuned over several gigahertz. Driving the system near its resonance with a strong pump tone enables amplification of a weak signal nearby in frequency. The gain of this amplification can be traded off with the bandwidth over which it occurs by adjusting the pump power and frequency. 

When the signal of interest is centered about the pump tone, the amplification process is noiseless (insofar as the JPA's internal loss is negligible) but phase-sensitive. In this mode of operation, the signal quadrature in phase with the pump is amplified by $\sqrt{G}$, while the one 90 degrees out of phase is squeezed by $1/\sqrt{G}$, where $G$ is the single-quadrature power gain. When detuned completely to one side of the pump, the signal is amplified independent of its phase, but the JPA's intermodulation gain implies that an extra noise term enters from the image frequency, symmetric with the signal about the pump. In this configuration the added noise of a lossless JPA is equal to the thermal noise at the image frequency, giving rise to the quantum limit $N_A=1/2$ at zero temperature.

It is important to note that operating in the phase-sensitive mode by itself does not eliminate the half-photon of noise associated with the zero-point motion of the blackbody gas in the cavity (i.e., the second additive term in Eq.~\eqref{eq:noise}). Moreover, the factor of two improvement in noise temperature for a lossless JPA in the phase-sensitive mode is exactly canceled by the loss of information in one quadrature. Thus, phase-sensitive operation offers no improvement in axion search sensitivity unless we can also eliminate the zero-point motion of the input noise, e.g., by initializing the cavity in a squeezed state \cite{zheng2016}. R\&D with this aim is ongoing within the collaboration; for the present, we operate the JPA in the phase-insensitive mode.

The JPA currently installed in the experiment has a maximum resonant frequency of 6.5~GHz and a gain bandwidth product $\sqrt{G}B = 26$~MHz. Half a flux quantum threaded through the area of a single SQUID loop in the array tunes the circuit through its full 2~GHz range; this extreme flux sensitivity necessitates specialized magnetic shielding, detailed in section~\ref{sub:shielding}. The circuit was fabricated using a Nb/Al-Ox/Nb trilayer process; the nonlinear inductance comes from 20 SQUIDs in series, where each SQUID comprises two Josephson junctions of critical current 6~$\mu$A in parallel. The flux bias is delivered through a coil comprising about 20 turns of superconducting wire wound around the copper box housing the JPA chip.

The JPA can be turned into a passive mirror very simply by tuning off the pump tone and tuning the resonant frequency $\sim$10 linewidths from the frequency of interest via the flux bias; we use this process to calibrate the absolute JPA gain. We set a gain target of $\simeq21$~dB, below the onset of bifurcation inferred from deviations from a gaussian distribution in the JPA's quadrature noise spectra, but large enough to overwhelm the $\sim20$~quanta added noise of the HEMT amplifier which follows the JPA in the receiver chain. At this operating point the JPA bandwidth is about 2.3~MHz. 

\begin{figure}[h]
\centering\includegraphics[width=0.45\textwidth]{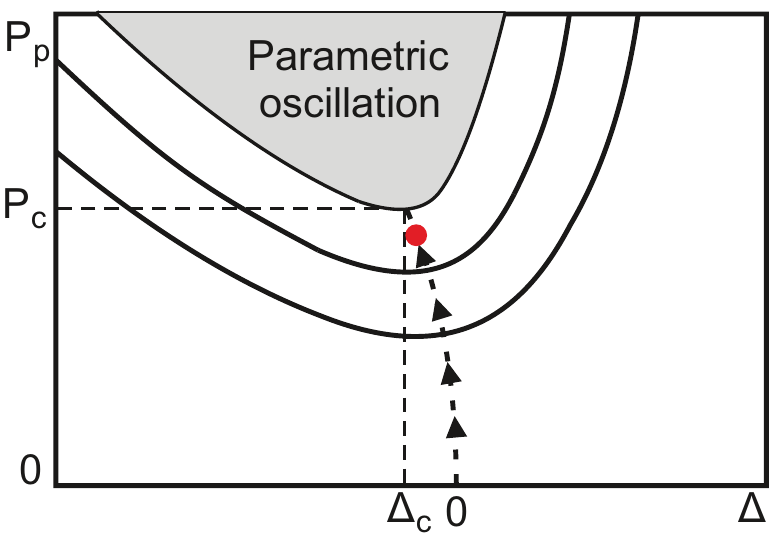}
\caption{Schematic representation of the parameter space for JPA biasing. All features are intended for illustrative purposes only; they do not represent real measurements or calculations. The pump power is on the vertical axis and the detuning $\Delta$ between the pump frequency and the 0-power $LC$ resonance of the JPA circuit is on the horizontal axis. The black curves are contours of constant gain, and the black arrows represent the path taken by the bias procedure outlined in the text, intersecting each gain curve at the minimum-power point. Beyond the critical point $(\Delta_c,P_c)$, the system begins to oscillate and can be bistable or multistable. We operate at lower power (red dot) to avoid this region of parameter space.\label{fig:jpa_biasing}}
\end{figure}

The procedure used to bias the JPA to this target gain is illustrated schematically in Fig.~\ref{fig:jpa_biasing}. The detuning $\Delta$ between the $LC$ resonance and the pump frequency is adjusted by varying the flux bias. The gain is optimized with respect to detuning at constant pump power, and the pump power is increased if the gain remains too low after this optimization; the pump frequency remains fixed throughout the biasing process. With this procedure we always obtain within 0.2~dB of the target gain, and always operate at the point of lowest pump power for a given gain, where the JPA is most stable. The biasing procedure is fully automated and incorporated into the LabVIEW code that controls the data acquisition (see section~\ref{sub:DAQ}). It typically takes $\sim6$~s to adjust the bias parameters after each cavity tuning step. 

Maintaining high gain throughout the tuning range requires the pump power and flux bias to be controlled to $0.01$~dB and 2 parts in $10^5$, respectively. The latter condition corresponds to 300~nA current resolution with our present bias coil, which we easily obtain using a 20-bit ADC with $10~\mu$V resolution and a homemade current source with 1 mA/V transconductance.

\subsection{Magnetic Shielding\label{sub:shielding}}

\begin{figure*}[h]
\centerline{\includegraphics[width=.8\textwidth]{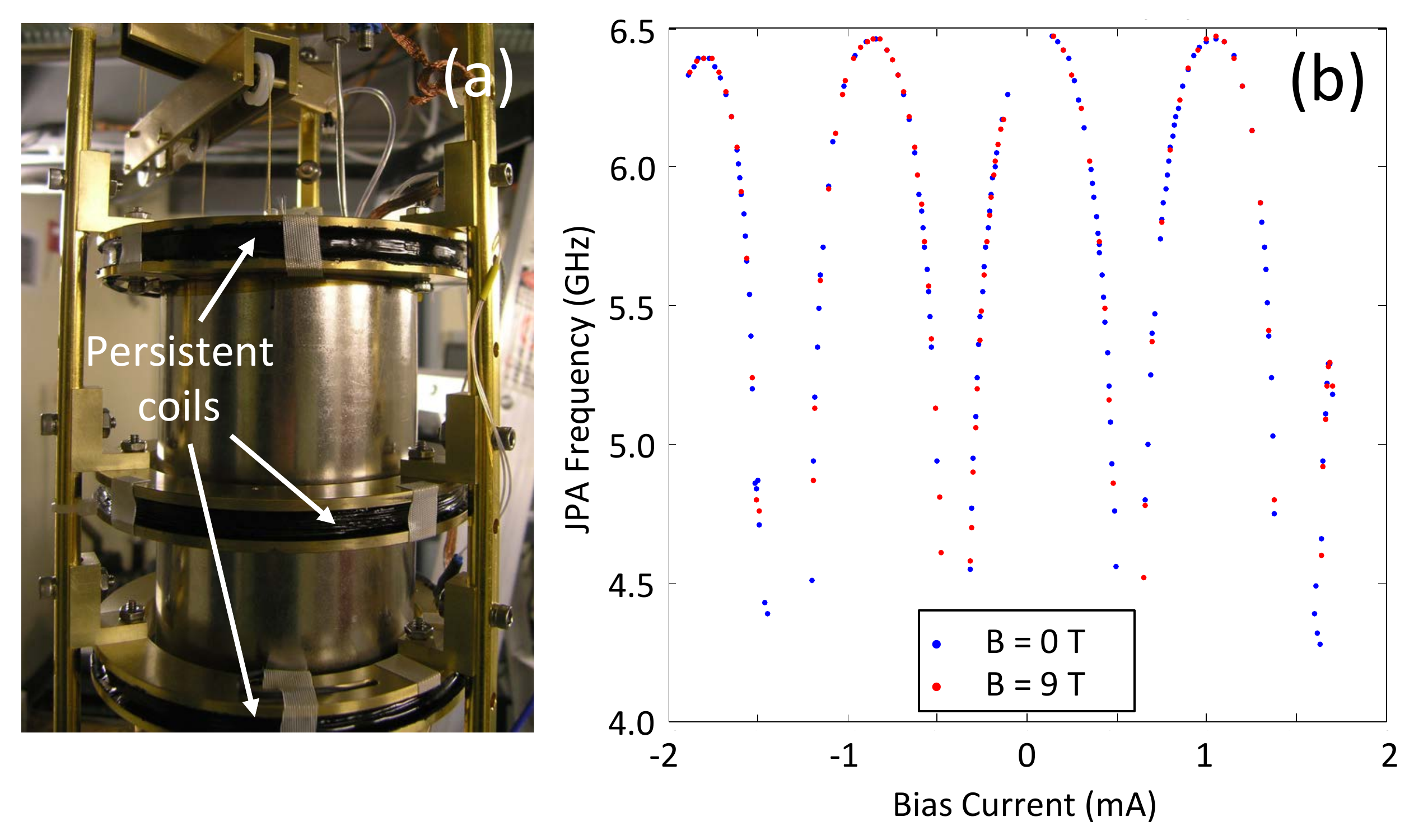}}
\caption{(a) Magnetic shielding of the JPA, including three superconducting persistent coils and (b) JPA frequency vs. flux bias current, showing no change in frequency as the main magnetic field is ramped from $0 - 9$~T. The residual flux through the JPA is inferred to be $\Delta \Phi \simeq 0.001$ flux quantum.}\label{Fig4}
\label{fig:flux_quanta}
\end{figure*}

Shielding of the sensitive receiver components, in particular the JPA, from the stray field of the 9~T magnet presents a complicated engineering problem. Also of concern are magnetic gradients at the location of the JPA that lead to excess amplifier noise due to gain fluctuations in the presence of mechanical vibrations. We have successfully attained the degree of magnetic shielding required for stable JPA operation, as shown in Fig.~\ref{fig:flux_quanta}(b).  

The outer elements of the JPA shield can be seen in Fig.~\ref{fig:flux_quanta}(a). The first level of shielding is from the bucking coils described in section~\ref{sec_magnet}. The next level of shielding comprises three persistent superconducting coils that provide a quasi-active shield against flux changes; given that they become superconducting before the 9~T field is ramped up, the flux inside will be maintained at the original ambient level. The axial spacing of the coils is not uniform because the bucked field gradient is asymmetric, with slightly higher flux at the lower end of the shield compared to the upper. The axial center of the shield is located at the field minimum.   

The superconducting shield coils each comprise 100 turns of Supercon Inc. SC-T48B-M-0.7mm Cu-clad NbTi SC wire, with 0.43~mm diameter NbTi core (nominal), 0.70~mm diameter Cu cladding, and 0.75~mm diameter HML (Polyimide) insulation. The windings are on 9.3 cm ID brass hoop forms and potted with Stycast 2850~FT. After winding, the coil wire ends are trimmed to about 15 cm, and the last 5 cm of wire are annealed by heating to red heat using a MAPP gas torch; this also burns off the Polyimide insulation. The heated regions are cleaned and smoothed using 400~grit alumina emery paper. The wire ends were bonded \cite{shielding1} to make a persistent superconducting connection by use of a Koldweld KBM-9 wire coldwelding machine; the dies we had available were too large, so the outer copper cladding was built up through acid copper electroplating. After electroplating, the ends were trimmed back 1.6 mm, and the wire ends bonded. A micrograph analysis of an axially mechanically sectioned bond shows that the NbTi material flowed together at the center and pushed the copper uniformly in a radial direction from the center.  

A similarly wound coil, with 24 cm OD, is clamped to the bottom of the DR mixing chamber plate. The fringe field of the 9~T magnet is about 300~G in this region; without the coil this is large enough to affect operation of the shielded cryogenic circulators (see section~\ref{sub:receiver}).  

The ferromagnetic shield comprises a nested pair of cylinders made of 1.5 mm thick Amumetal 4~K, annealed per Amuneal's proprietary process. The bottoms of the can have welded-on disks, while the tops are closed with tight-fitting lids that overlap 1.25 cm of the cylinder walls. The outer shield (FS1) has length 15.2 cm and OD 8.9 cm, while the inner shield (FS2) has length 13.3 cm and OD 7.6 cm. Both top lids have holes for the tuning rod (1.25 cm) and also for the JPA thermal link and transmission line. There are also three small holes, placed at 120$^\circ$ in both the top and bottom, that allow Kevlar tuning lines to pass through both shields.  

Between the two ferromagnetic shields is a Pb superconducting shield, fabricated from 1.6 mm Pb sheet, which fits closely on the outside of FS2 and has a removable close-fitting lid. The Pb shield can is about 12.0 cm long, and covers FS2 to just below the lid. The Pb lid is about 3.8 cm long, giving an overlap with the cylinder of 2.5 cm. The inside of the Pb can is coated with clear acrylic paint to avoid electrical contact and thermal currents, and is glued to FS2 using Loctite~680.  

The final passive shield component is a 0.13 mm~thick Nb sheet (99.8~\% purity, Alfa-Aesar) that lines the inner surface of FS2. This sheet forms an ``open cylinder'' in that it has no endcaps, and there is an effective axial slit along its length which allows flux that would otherwise be trapped to escape. The sheet is long enough so that the ends overlap by about 1.25 cm, and the sheet is coated with acrylic paint, again to prevent the formation of electrical contacts. The sheet is glued to the inner surface using Loctite. This sheet enforces a boundary condition that the magnetic field must be axially homogeneous along its surface. The ferromagnetic boundary condition at the bottom and at the lid of FS2 is that the magnetic field is perpendicular to the surface, which matches perfectly the boundary condition imposed by the Nb sheet. The combined effect of the superconducting and ferromagnetic boundary conditions is to substantially reduce field gradients at the JPA.

A G-10 disk and nylon standoffs are used to support the JPA at the axial center of the shield, with a radial offset to allow the tuning rod to pass. The standoffs are screwed to 8-32 studs that extend from OFHC Cu spacers between the two ferromagnetic shields; the cooling of the shield assembly is through these studs and spacers which clamp the shields to each other and to a Cu plate affixed to the gantry. The JPA itself is thermally linked to the gantry via a Cu braid that exits through the top of the FS2/Pb shield assembly.

A final active layer of magnetic shielding is provided by feedback to the JPA flux bias coil, discussed in section~\ref{sub:receiver}. 

\subsection{Receiver and microwave measurements\label{sub:receiver}}

The cryogenic microwave layout and room-temperature microwave/IF layout are depicted schematically in Fig.~\ref{fig:cryo_setup} and Fig.~\ref{fig:RT_setup} in the appendix, respectively. The receiver signal path is shown in blue in both diagrams. This section will describe the microwave portion of the receiver, the RF input lines, and other detector functionality which is best understood in reference to Fig.~\ref{fig:cryo_setup} and Fig.~\ref{fig:RT_setup}. The IF part of the receiver chain is discussed in section~\ref{sub:DAQ} below.

\begin{figure}[h]
\centering\includegraphics[scale=0.55]{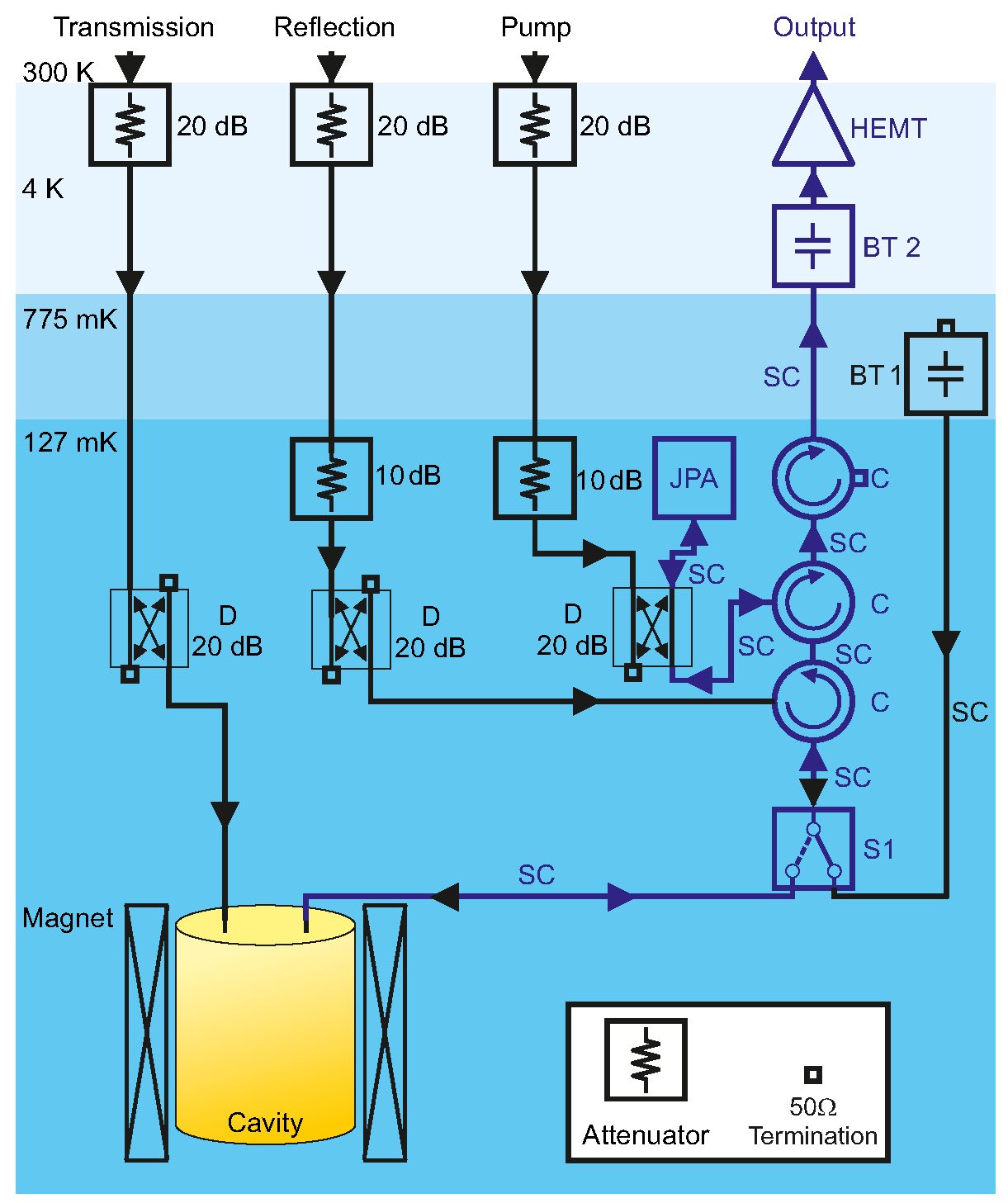}
\caption{\label{fig:cryo_setup} The cryogenic microwave layout. Blue arrows indicate the receiver signal path from the cavity to room temperature; black arrows indicate other paths used for network analysis, noise calibration, and JPA biasing. Component part numbers and manufacturers are listed in Table~\ref{tab:parts}, in the Appendix.}
\end{figure}

The input of the receiver is the cryogenic microwave switch S1, which may be toggled to connect the receiver to either the cavity or a termination thermally anchored to the DR still plate at $T_H=775$~mK. This switch allows us to calibrate the receiver's added noise \textit{in situ} via a procedure described in detail in section~\ref{sub:noise}.

The JPA, playing the role of the preamplifier, comprises the magnetically shielded JPA circuit described in section~\ref{sub:jpa}, a directional coupler for the pump tone input, and a commercial ferrite circulator to separate input and output signals. Two other circulators are required to isolate the JPA from both the backaction of the second-stage amplifier and its own backaction in reflection from the cavity. Signals exiting the JPA are amplified further by a HEMT amplifier at 4~K and another low-noise transistor amplifier at room temperature, and then routed both to a commercial vector network analyzer (VNA) and to an IQ mixer (M2) serving as the input of a homemade spectrum analyzer. The former path is used for cavity and receiver characterization, the latter for periodic noise calibrations and the cavity noise measurements constituting the axion search dataset.

Swept tones produced by the VNA may be directed via software-controlled switches S2 to any of three fridge input lines, through which they are transmitted through the cavity, reflected off the cavity, or sent directly to the JPA, bypassing the cavity. The first two lines are used to measure the cavity parameters $\nu_c$, $Q_L$, and $\beta$; the third line is used for JPA biasing and gain measurements, as discussed in section~\ref{sub:jpa}. Room-temperature attenuators were chosen to equalize sweep power incident on the JPA through each path. Attenuators at 4~K and base temperature reduce room-temperature thermal noise and the phase noise of the JPA pump generator to $\sim$~mK contributions we can safely ignore. 

$0.085''$~NbTi/NbTi coaxial cables are used in the receiver signal path between base temperature and 4~K; these cables (marked SC in Fig.~\ref{fig:cryo_setup}) remain superconducting in the 9~T field due to flux pinning. Stainless $0.085''$~coax is used between room temperature and 4~K in all four lines and down to base temperature in the input lines. All four coaxial lines are thermalized at each stage of the fridge with gold-plated copper clamps; the output line inner connectors are thermalized via a bias tee at 4~K with its DC input shorted to ground. NiCr resistors are used in all cryogenic attenuators and terminations.

The room-temperature microwave chain includes three signal generators from Keysight in addition to the VNA. An 8340B serves as the local oscillator (LO) for both the spectrum analyzer system and the flux feedback system discussed below. An E8257D with an ultra-low phase noise option ($-120$~dBc/Hz at 100 kHz detuning from the carrier) provides the pump tone for the JPA. Finally an N5183B is used to inject synthetic axion-like signals into the cavity transmission line. All three generators, along with the VNA and the digitizer board used in the spectrum analyzer, lock to a common 10~MHz reference provided by a Stanford Research Systems (SRS) FS725m rubidium source (not pictured).

\begin{figure}[h]
\centering\includegraphics[width=0.45\textwidth]{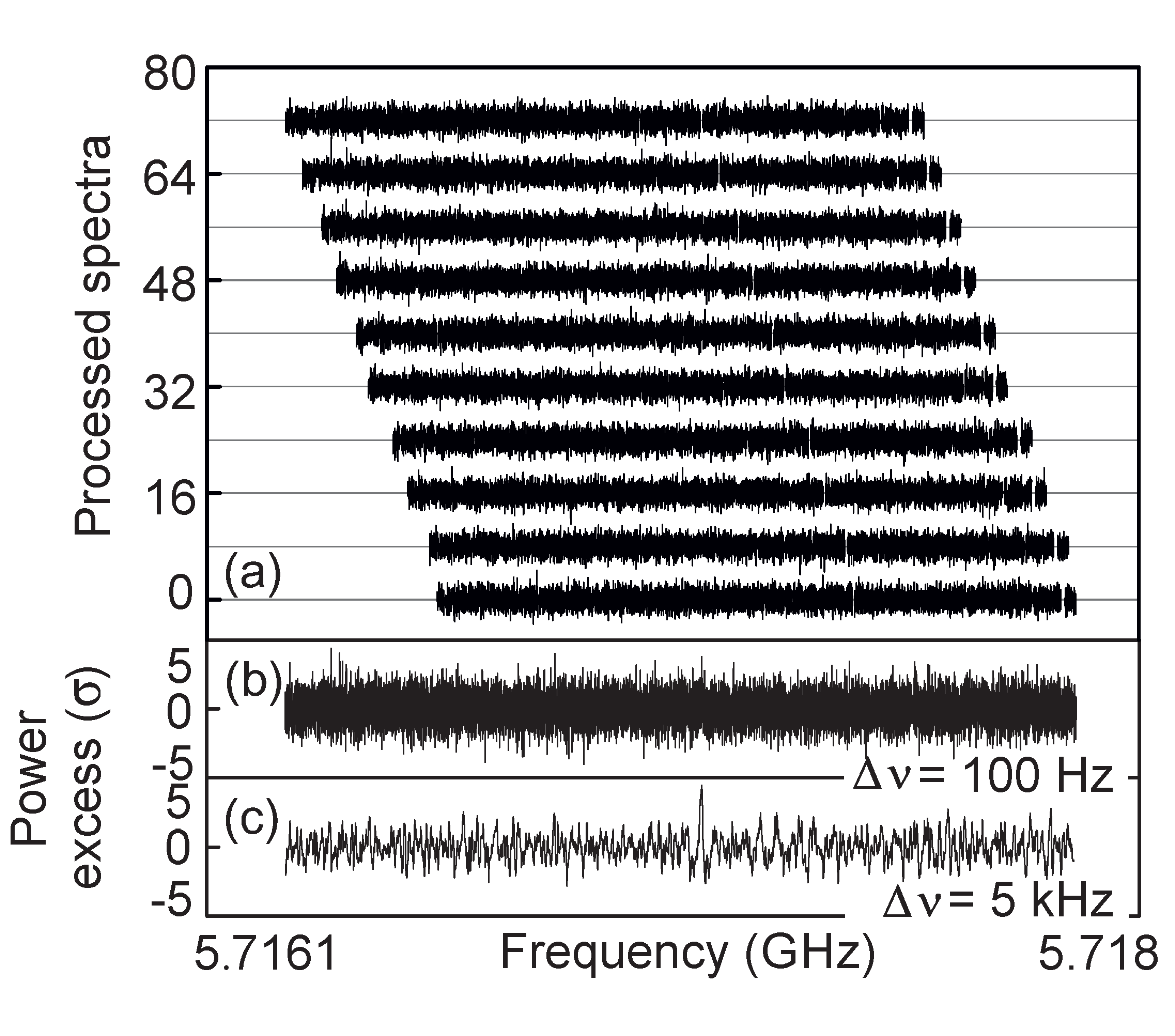}
\caption{\label{fig:spectra} (a) 100~Hz-resolution power spectra from individual 15-minute integrations around a frequency at which a synthetic axion signal was injected. (b) The optimally weighted combined spectrum, still at 100 Hz resolution. (c) The combined spectrum after rebinning to 5~kHz with weighting that takes into account the axion lineshape; the synthetic axion is clearly visible. See~\cite{PRL2016} for an overview of the analysis procedure, which will also be described more thoroughly in a forthcoming publication.}
\end{figure}

To generate synthetic axion signals with the expected linewidth ($\sim$5~kHz for a $25~\mu$eV axion), we inject band-limited white noise into the FM port of the N5183B; the linewidth is then controlled by the modulation depth. The synthetic axion power in the cavity is only known to $\pm3$~dB due to unknown cryogenic insertion losses of individual components. This uncertainty prevents us from using the synthetic axion system to precisely calibrate our sensitivity, but it is still useful as a way to validate both the data acquisition and analysis procedures. Our data acquisition software includes functionality for blind injection of axion signals at random frequencies throughout a data run; Fig.~\ref{fig:spectra} illustrates one such injection in our first run, with a nominal intracavity power level of $-190$~dBm.

The room-temperature signal paths used exclusively by the JPA flux feedback system mentioned at the end of section~\ref{sub:shielding} are shown in pink in Fig.~\ref{fig:RT_setup}. The JPA flux bias is modulated at 26~Hz by adding an oscillating current (controlled by an Agilent 33220A function generator) to the DC bias current set in software. The modulation frequency is limited by eddy current shielding in the Cu JPA enclosure; the amplitude is set to yield a modulation depth of $\sim0.1$ JPA linewidth. When the DC flux bias is offset from the value that maximizes the JPA gain at fixed pump power and frequency, the JPA gain is modulated at the same frequency as the flux with phase determined by the sign of the offset. When the bias modulation is centered on the optimal value, only higher harmonics are present in the gain variation. Thus the JPA gain variation at the modulation frequency may be used as an error signal in an analog feedback system to stabilize the flux bias.

During noise measurements we use the VNA to inject a weak CW probe tone through the pump line at 30~kHz detuning from the pump. The function of the feedback circuitry in the upper left corner of Fig.~\ref{fig:RT_setup} is to provide an LO at the appropriate IF frequency to bring the probe tone to DC on the IF side of the mixers M3. Both quadrature outputs are amplified and squared using AD633 analog multipliers; the sum of squared signals is a measure of the received power in the probe tone and thus of the JPA gain. A Stanford Research Systems SR510 lock-in amplifier measures the variation of the probe tone power at the modulation frequency. By adding the SR510 output to the bias current we obtain a simple proportional feedback loop; the SR510 gain, phase, and filtering are chosen to provide a stable feedback signal. During JPA biasing and sweep measurements the probe tone is not present because the VNA is otherwise occupied, so both modulation and feedback are interrupted by switching off the signal and reference outputs of the 33220A in software.

\subsection{Data acquisition and operations\label{sub:DAQ}}
The spectrum analyzer comprises the discrete components between M2 and ADC in Fig.~\ref{fig:RT_setup}, along with LabVIEW code that computes FFTs, implements image rejection, and averages power spectra in parallel with timestream data acquisition. During all noise measurements the LO for M2 is set 780~kHz above the TM$_{010}$ mode frequency, the JPA pump is set 820~kHz below the mode, and both IF channels are sampled by the GaGe Oscar CSE4344 PCIe digitizer board at 25~MS/s. 

Each IF channel is sensitive not only to the RF frequencies of interest in the lower sideband of the LO but also to unwanted image frequencies in the upper sideband. The 90$^\circ$ relative phase shift between the two otherwise identical IF outputs of an IQ mixer may be exploited for image rejection: adding the Q output to the I output with a $+(-)90^\circ$ phase shift suppresses the upper (lower) LO sideband. This is the operating principle of commercial image reject mixers, which only work at particular IF frequencies because they implement the required $90^\circ$ phase shift in hardware.

We implement image rejection in the frequency domain in software: this scheme works at any IF frequency, with the only possible drawback being that amplitude and phase mismatches in the discrete IF components in the two channels can limit the degree of image rejection. By taking FFTs of both the I and Q channels and defining $X(\omega) = \left(\text{Re}[I(\omega)] - \text{Im}[Q(\omega)]\right) + i\left(\text{Im}[I(\omega)] + \text{Re}[Q(\omega)]\right)$ we obtain rejection of the upper sideband better than 20 dB at all IF frequencies of interest in the power spectrum $\left|X(\omega)\right|^2$.

More specifically, 14-bit ADCs on the GaGe board digitize both channels simultaneously in 5~s segments, then transfer each segment to PC RAM. The total data in each segment across both IF channels is 437.5 MB, and the time required to transfer this data to RAM is 1.2~s, which caps the data acquisition efficiency at 80\%. The \textit{in situ} processing code divides each segment into 500 non-overlapping 10~ms records in each channel, computes the FFT of each record with no windowing, combines the I and Q FFTs corresponding to the same 10~ms time slice to implement image rejection, constructs a power spectrum from each sample of $X(\omega)$, and averages all 500 power spectra corresponding to each segment. All processing for each 5~s segment occurs in RAM in parallel with the acquisition of the next segment. At the end of the data acquisition period (typically 15 minutes), the power spectra from all segments are averaged together to obtain a single spectrum obtained from $100~\text{Hz}\times15~\text{minutes}=9\times10^4$ averages, which is written to disk.

Thus the output at each tuning step is a heavily averaged power spectrum with 100 Hz resolution extending from DC to the 12.5~MHz Nyquist frequency. The usable IF bandwidth extends to $\sim2.5$~MHz, limited by the bandwidth of the low-pass filters F2. The sampling rate is set so far above this to eliminate the need for a high-order filter with significant passband ripple on the output of the IF amps A2. The function of the F1 filters is to attenuate RF leakage into the IF chain.

The 100 Hz spectral resolution is much smaller than the $\sim5$~kHz expected linewidth of a virialized axion signal -- thus data from this detector can be used to set more stringent limits on the abundance of non-virialized axions with $Q_a \lesssim 10^7$ without any additional hardware. A technical advantage of the narrow resolution is that it enables us to better identify and eliminate individual IF bins contaminated by narrowband interference. Empirically some of these ``IF spikes'' are associated with ground loops and others are due to room-temperature electronics. The DC blocks and baluns in Fig.~\ref{fig:RT_setup} were added in commissioning to eliminate as many of the IF spikes as possible; remaining spikes are flagged and removed as part of the analysis procedure.

\begin{figure}[h]
\centering\includegraphics[width=.5\textwidth]{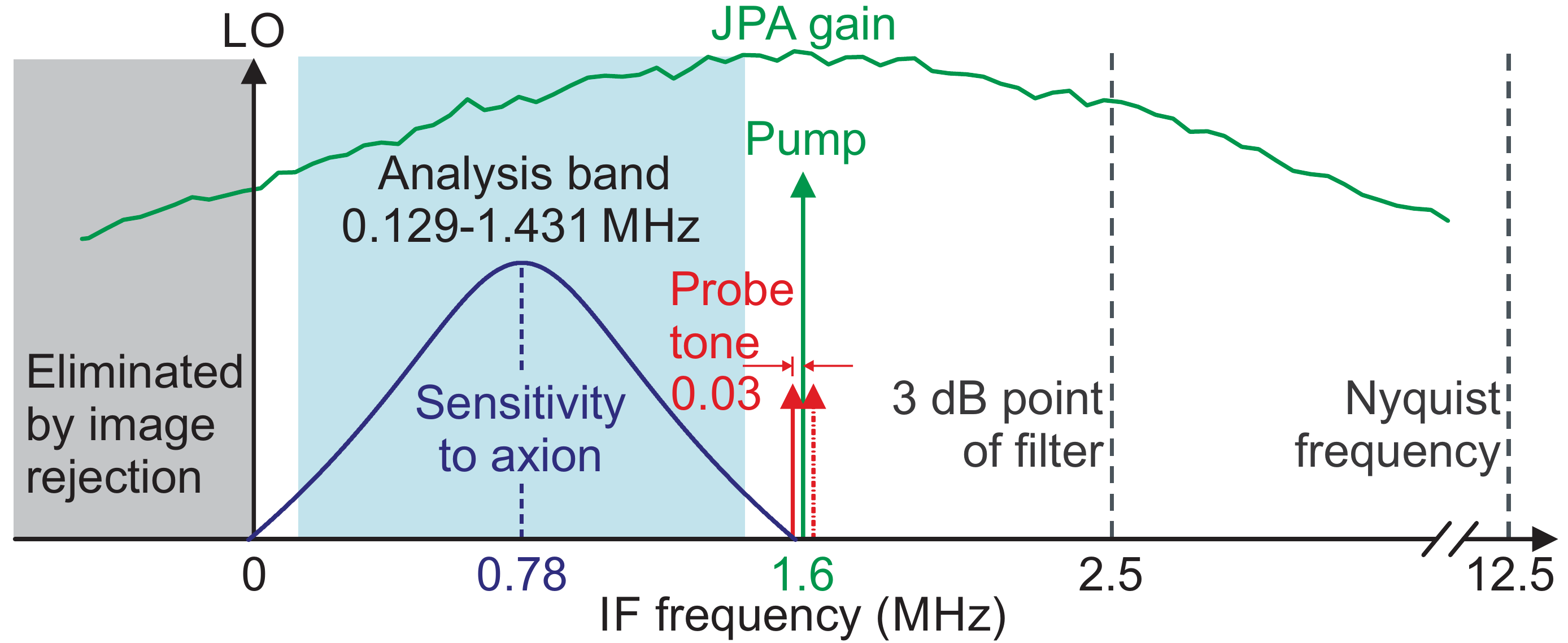}
\caption{Diagram illustrating the IF setup for the experiment. Note that the IF frequency axis is reversed relative to the RF frequencies of these features. The JPA gain profile and TM$_{010}$ Lorentzian profile are plotted using real data and a fit to real data, respectively. Both plots have logarithmic y axes; the absolute scale of the sensitivity plot is arbitrary. The red dot-dashed arrow indicates the probe tone created on the opposite side of the pump by the JPA's intermodulation gain. Images of the axion-sensitive Fourier components around the cavity, created on the other side of the pump by the same process, are omitted in the diagram for clarity. \label{fig:IF_layout}}
\end{figure}

Fig.~\ref{fig:IF_layout} is a simplified diagram of the IF setup showing the most relevant features discussed above, and illustrating that we further limit our analysis to 1.302~MHz centered on the cavity. The width of this analysis band is roughly twice the maximum cavity linewidth; axion conversion at larger detunings from the cavity mode frequency contributes negligibly to the SNR. The LO frequency must lie outside the analysis band to avoid superimposing different analysis band Fourier components, and the pump frequency must lie outside the analysis band for the JPA to operate in the desired phase-insensitive mode (see section~\ref{sub:jpa}). The LO and pump frequencies also cannot be equal, or else the I and Q channels will pick out (not necessarily equal) linear combinations of the JPA's amplified and squeezed quadratures, and image rejection will fail. 

These constraints indicate that the analysis band should be centered roughly halfway between the LO and pump in each spectrum. If the analysis band were too close to the pump, the spectrum would be contaminated by both the feedback probe tone and the pump tone's phase noise. If the analysis band were too close to DC, $1/f$ noise would dominate, and the relative contribution of the 4~K HEMT amplifier to the receiver's added noise would grow with detuning from the pump frequency. The precise positioning of the analysis band was tweaked to exclude bins in which IF interference was most persistent.

Data acquisition for the experiment is fully automated and controlled by a LabVIEW program. At the beginning of each iteration, the DAQ program tunes the TM$_{010}$ mode, measures $\nu_c$ and $Q_L$ with a VNA sweep through the cavity transmission line, then sets the LO and pump frequencies based on the new value of $\nu_c$. It optimizes the JPA gain as described in section~\ref{sub:jpa}, measures the gain profile over 5~MHz centered on the pump, turns on the flux feedback system, and makes the 15-minute power spectrum measurement described above. At the end of the power spectrum measurement, the JPA gain and cavity transmission are each measured again, to identify drifts of the cavity mode associated with mechanical relaxation (see section~\ref{sub:cavity}) and unusually large bias flux jumps that the feedback system was unable to correct for; roughly 0.6\% of spectra were cut from our initial analysis due to anomalous flux or frequency drifts. Finally, $\beta$ is measured with a reflection sweep. Noise calibrations (see section~\ref{sub:noise}) are also performed intermittently to constrain variation of $T_S$ throughout the run. The average live time efficiency during the run is 72\%.

All sweep data is saved to disk along with the 100~Hz averaged power spectrum and critical parameter values such as the LO frequency; this amounts to about 3~MB per iteration. Data is stored locally at Yale and also transferred to a remote server at Berkeley for back-up and long term storage. The offline analysis procedure is outside of the scope of this paper, and will be the subject of a forthcoming publication.

\subsection{Noise calibration}
\label{sub:noise}

A simple way to measure the added noise of any microwave receiver is via a Y-factor measurement, in which we connect the receiver input to two 50~$\Omega$ loads at known temperatures $T_C$ and $T_H$, and measure the hot/cold noise power ratio

\begin{equation}
Y = \frac{P_H}{P_C} = \frac{N_H + N_A}{N_C + N_A},
\label{eq:yfactor}
\end{equation}

\noindent where $N_{C} (N_H)$ is the thermal noise from the cold (hot) load including the zero-point contribution, and $N_A$ is the receiver's added noise. We can then solve Eq.~\eqref{eq:yfactor} for $N_A$ to obtain

\begin{equation}
N_A = \frac{N_H - YN_C}{Y - 1}.
\label{eq:added}
\end{equation}

In our experiment, $N_H$ comes from a 775~mK termination, and $N_C$ comes from the cavity (on resonance) or from a terminated port on the directional coupler in the reflection input line, reflected off the cavity (off resonance; see Fig.~\ref{fig:cryo_setup}). $N_A$ includes the added noise of the JPA preamplifier and also the added noise of subsequent amplifiers referred to the JPA input; of those, in practice only the 4~K HEMT contribution is non-negligible.

Eq.~\eqref{eq:SNR} indicates that the effect of imperfect power transmission efficiency $\eta$ between the cavity and the JPA is to rescale the total noise by $1/\eta$. It is convenient to define $\eta=\eta_0\eta_1$, with $\eta_0$ between the cavity and S1 and $\eta_1$ between S1 and the JPA input. Then $N_A$ measured by the Y-factor method will also include an additional additive term

\begin{equation}
N_\eta = \frac{1-\eta_1}{\eta_1}\left(N_C + N_A'\right),
\label{addterm}
\end{equation}

\noindent where $N_A'$ indicates the sum of amplifier contributions. This term arises because the lossy elements that contribute to $\eta_1$ are at temperature $T_C$ (see Fig.~\ref{fig:cryo_setup}), and thus attenuate both the hot load noise and the axion signal but not the cold load noise. To derive Eq.~\eqref{addterm} more formally it is important to note that the lossy elements also generate thermal noise given by $(1-\eta_1)N_C$. Thus our measurement of $N_A$ yields only the net effect of the JPA added noise, HEMT noise, and $\eta_1$, not any of these components individually. The effect of $\eta_0$ on the SNR cannot be measured \textit{in situ}; we estimate 0.6~dB from the long superconducting cable and connector losses, and take this loss into account in calculating our exclusion limits.

The general expression in Eq.~\eqref{eq:added} also must be modified to account for several specific features of our receiver. First, as noted in section~\ref{sub:jpa}, in our present configuration the JPA's added noise is simply the thermal noise on the opposite side of the pump from the cavity, coupled into the analysis band via image gain. Thus, the JPA contribution to the added noise differs by a known amount in the two switch configurations: if we define $N_A$ to be the added noise in cold load measurements, the added noise with the switch pointed at the hot load is $N_A + N_H - N_C$, since the loss and HEMT contributions do not change with temperature. Second, both sides of Eq.~\eqref{eq:added} are in principle functions of IF frequency within any given measurement; in particular the lower JPA gain at the low-IF-frequency end of the analysis band (see Fig.~\ref{fig:IF_layout}) implies less suppression of the HEMT noise and thus larger $N_A$. Third, actuating the switch affects the standing wave pattern due to slight impedance mismatches on the cryogenic transmission lines, resulting in a small change in the pump power delivered to the JPA. We rebias the JPA every time we actuate the switch, but small changes in the JPA gain and bandwidth between switch configurations are still possible. We take all three of these effects into account by generalizing Eq.~\eqref{eq:added} to

\begin{equation}
N_A(\nu) = \frac{2\left[A(\nu)N_H - Y(\nu)N_C\right]}{Y(\nu) - A(\nu)} + N_C,
\label{eq:added_new}
\end{equation}

\noindent where $A(\nu)=G_H(\nu)/G_C(\nu)$ is obtained from the measured hot/cold JPA gain profiles. 

During commissioning of the experiment, we consistently obtained $N_A\simeq1.35$~quanta from Y-factor measurements far detuned from the cavity mode independent of RF frequency. This result is consistent with the added noise of an ideal nondegenerate JPA at $T_C$ (0.63~quanta), $\simeq0.2$~quanta input-referred HEMT noise, and an additional $\simeq0.5$~quanta which we can attribute to $\simeq2$~dB loss before the JPA. This is a plausible value result for $\eta_1$, if a little on the high side. 

In cold load noise measurements near the TM$_{010}$ frequency we observe an additional Lorentzian excess $N_{\text{cav}}(\nu)$ centered on the IF frequency of the mode. Naively applying Eq.~\eqref{eq:added_new} to such measurements would treat the $N_{\text{cav}}$ as a contribution to the added noise (because this expression assumes the input noise in the cold load is completely characterized by the single number $T_C$), and therefore overestimate the effect on the total noise $N_{\text{sys}}$, since $N_A$ appears in both the numerator and denominator of Eq.~\eqref{eq:yfactor}, whereas $N_C$ appears only in the denominator.

Therefore, we make the ansatz that the added noise is the same on-resonance as off-resonance, which allows us to quantify the peak excess contribution $N_{\text{cav}}(\nu_c) \simeq1$~quantum. During detector commissioning we examined the dependence of $N_{\text{cav}}$ on the mixing chamber plate temperature $T_C$, and observed that the excess vanished for $T_C \gtrsim 550$~mK. This behavior indicates that $N_{\text{cav}}$ has a thermal origin as opposed to being an artifact of feedback between the JPA and cavity mode. Studies of the dependence of $N_{\text{cav}}$ on JPA gain and on $\beta$ further disfavor any sort of feedback explanation, which validates the above ansatz. 

\begin{figure}[t]
\centering\includegraphics[width=.48\textwidth]{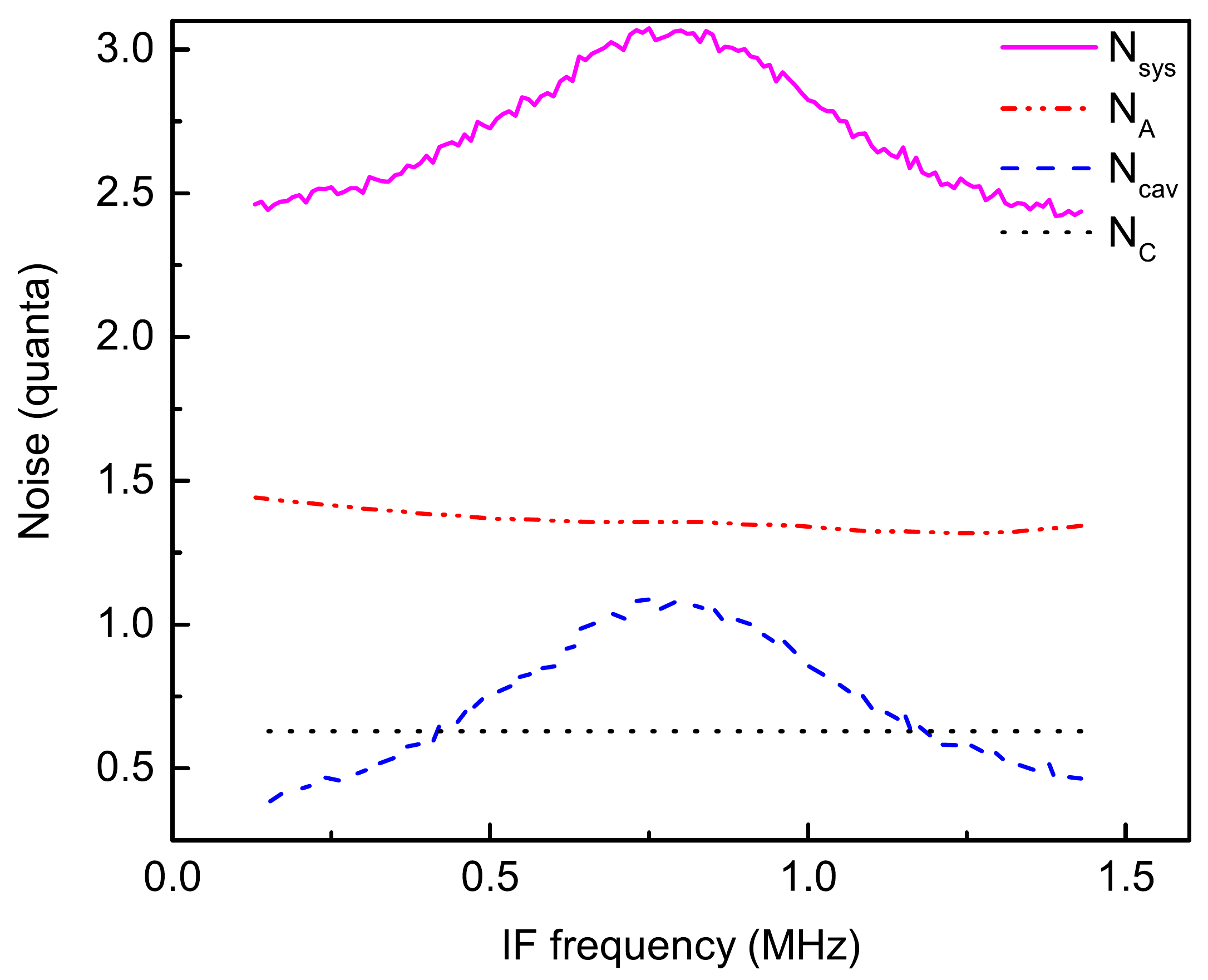}
\caption{A representative noise measurement. $N_C$ (black dotted line) is obtained from thermometry, $N_A$ (red dot-dashed line) is obtained from the average of off-resonance Y-factor measurements, and $N_{\text{cav}}$ (blue dashed line) is from a single Y-factor measurement during the data run. $N_{\text{sys}}$ (pink solid line) is the sum of these contributions.\label{fig:Yfactor}}
\end{figure}

Having established that $N_{\text{cav}}$ is of thermal origin, the most likely culprit is a poor thermal link between the tuning rod and the cavity barrel, as the only thermal connection in the present design is through a thin-walled alumina shaft. An improved thermal link through the axle of the tuning rod has been designed and is being implemented; it should bring the tuning rod into equilibrium with the mixing chamber and improve the linear coupling sensitivity by 20\% \footnote{Subsequent to the review of this paper, this thermal link has been implemented and the thermal effect has indeed been substantially mitigated.}. 

In practice, our \emph{in situ} noise calibration procedure consists of the following steps, repeated for each switch configuration: we bias up the JPA to target gain, measure the gain profile, and take 5 s of noise data, from which we construct a power spectrum with 10 kHz resolution, for a total of 130 data points in the analysis band. We wait 2 minutes for thermal transients to die down after actuating the switch in either direction; the heat due to each switch actuation is 10 mJ. 

From these measurements we obtain a typical total noise of $N_{\text{sys}}\simeq3$ quanta on resonance, falling to $\simeq2.2$ quanta at the edges of the analysis band. The overall fractional error $\delta N_{\text{sys}}\sim6\%$ is dominated by $\sim17\%$ uncertainty in $N_{\text{cav}}$; $\delta N_A$ is only $\sim 4\%$, and the change in $N_C$ is negligible even allowing for a $\pm20$~mK calibration error in the mixing chamber thermometry. Individual contributions to $N_{\text{sys}}$ are plotted in Fig.~\ref{fig:Yfactor} for a representative measurement. This is the lowest noise demonstrated to date in a microwave cavity axion search.

\section{Summary and Conclusions}

We have demonstrated \textit{in situ} near-quantum-limited noise performance in the first operation of this experiment, and achieved sensitivity to cosmologically relevant QCD axion models with $g_\gamma \geq 2.3\times g_\gamma^{\text{KSVZ}}$~\cite{PRL2016}. That sensitivity this close to the KSVZ line can be reached in an experiment of 1.5~L volume (c.f. the 200 L volume of the ADMX detector~\cite{Pen00,Asz11}) underscores the role of technology in the microwave cavity experiment. This platform is now poised to explore innovative concepts in both amplifiers and cavities to  increase sensitivity and decrease scan time. A receiver based on injecting a squeezed state of vacuum into the cavity by one JPA and reading it out with another will be the first such innovation to be explored \cite{zheng2016}. Beating the Standard Quantum Limit with such a receiver has been demonstrated on the bench~\cite{Mal11}, but this will be the opportunity to validate the feasibility of such a receiver in the rigors of an actual operating environment.  

On the cavity side, a photonic band gap resonator is being designed to eliminate all interfering TE modes. As the TM mode of interest is tuned in frequency, avoided crossings with TE modes block a significant fraction of frequency coverage. This loss of coverage has been the bane of rapid and efficient mass coverage to date. Such structures have been thoroughly explored in accelerator physics~\cite{Smi02,Nan13}; their successful adaptation to the microwave cavity axion experiment will largely pivot on an effective and reliable means of tunability. Similarly, cavities are being explored adapting design principles of distributed Bragg reflector structures, which attain much higher quality factors $Q$, providing greater sensitivity, and thereby scan rate, for the experiment~\cite{Flo97}.

\section{Acknowledgements}

This work was supported under the auspices of the National Science Foundation, under Grants PHY-1067242, and PHY-1306729, the Heising-Simons Foundation under Grants 2014-181, 2014-182, and 2014-183, and the U.S. Department of Energy by Lawrence Livermore National Security, LLC, Lawrence Livermore National Laboratory under Contract DE-AC52-07NA27344. We thankfully acknowledge the critical contributions by Matthias B\"uhler of Low Temperature Solutions UG for design of and upgrades to the cryogenic system. M.S. is supported by the National Science Foundation Graduate Research Fellowship Program under Grant no. DGE-1106400. 

\section*{}

\bibliography{simanovskaiaNIM2016arxiv}

\appendix

\onecolumn

\section{Room-Temperature Receiver Layout}

\begin{figure}[h]
\centering\includegraphics[scale=.6]{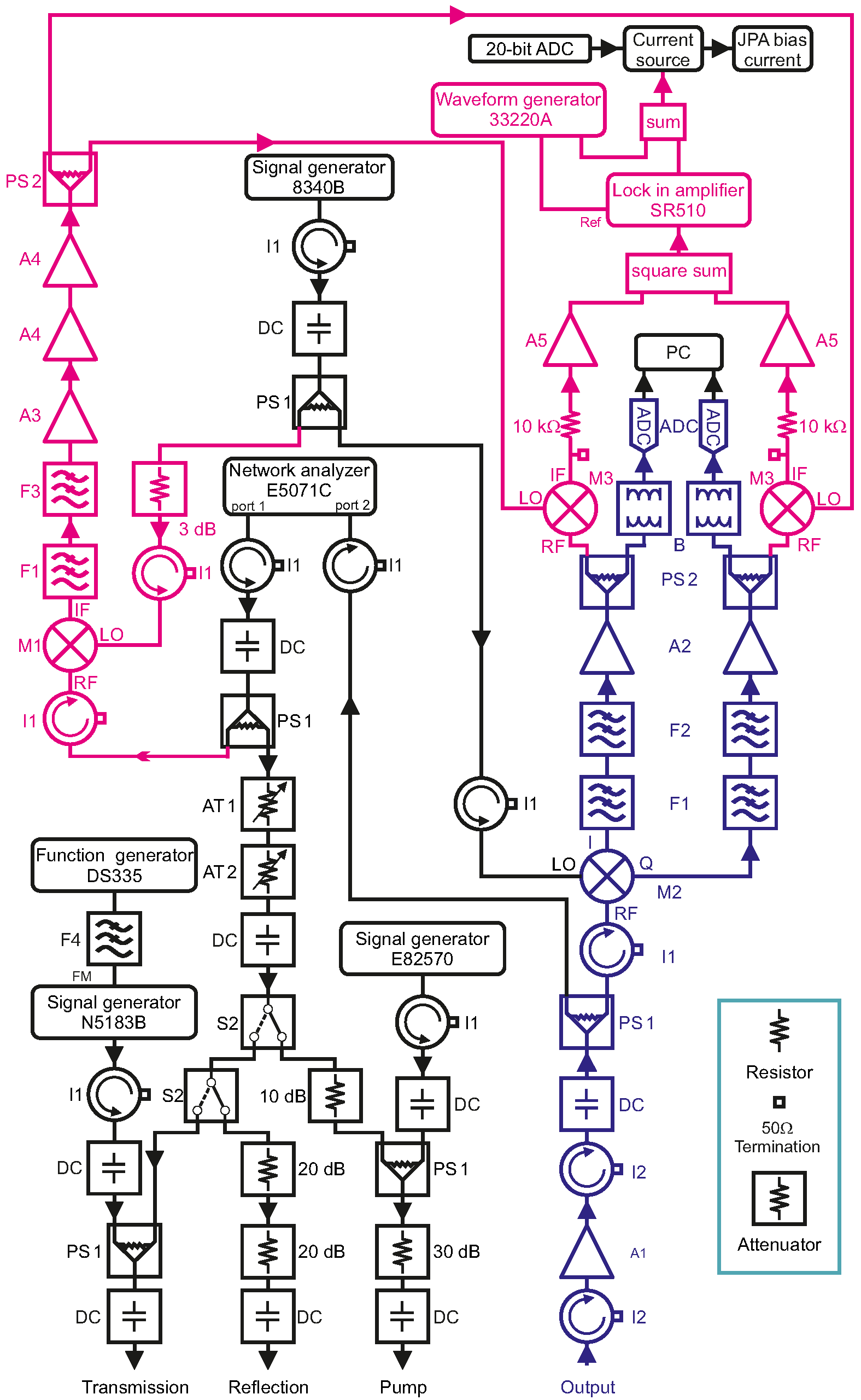}
\caption{\label{fig:RT_setup} The room-temperature microwave/IF layout. Blue arrows indicate the receiver signal path from the top of the fridge to the ADCs; black arrows indicate other paths used for network analysis, JPA biasing, providing LO power, and synthetic axion signal injection. Parts of the chain used exclusively by the JPA flux feedback system are indicated in pink. Component part numbers and manufacturers are listed in table~\ref{tab:parts}; those shown on the diagram are from Keysight or Stanford Research Systems.}
\end{figure}

\begin{table*}[h]
\caption{Component part numbers.\label{tab:parts}}
\begin{tabular}{llll}
\hline
\hline
  Thermal Environment & Label & Type & Supplier: Part \#\\
\hline
\hline
 Cryogenic & & & \\ 
  	& BT1 & Bias tee & Mini-Circuits: ZX85-12G+ (ferrite removed)\\
	& BT2 & Bias tee& Anritsu: K250\\
	& C & Circulator& Pamtech Inc.: CTH1184K18 \\
	& D & Directional coupler& Pasternack: PE2211-20\\
	& HEMT & HEMT amplifier& Low Noise Factory: LNF-LNC4\textunderscore8A\\
  	& S1& Switch & Radiall: R577443005\\
	& SC& NbTi/NbTi coax & Keycom: UPJ07\\
\hline
 Room-temperature & & & \\ 
  	& A1 & RF amplifier& Miteq: AMF-4F-04001200-15-10P\\
	& A2 & IF amplifier& Homemade: based on Fig.~2 in \cite{ca3018}\\
	& A3 & IF amplifier& Mini-Circuits: ZFL-500LN-BNC+ \\
	& A4 & IF amplifier& Stanford Research: SR445A\\
	& A5 & IF amplifier& Stanford Research: SR560\\
  	& AT1 & Step attenuator& Agilent Technologies: 8496H\\
	& AT2 & Step attenuator& Agilent Technologies: 8494H\\
	& B & Balun & North Hills: 0017CC \\
	& DC & DC block& Inmet: 8039\\
  	& F1 & Low-pass filter& Mini-Circuits: VLFX-80\\
	& F2 & Low-pass filter& Mini-Circuits: SLP-1.9+\\
	& F3 & Low-pass filter& Mini-Circuits: BLP-2.5+\\
	& F4 & Low-pass filter& Homemade: 3.39~kHz single pole RC filter\\
	& I1 & Isolator& Ditom Microwave: D314080\\
	& I2 & Double isolator& Ditom Microwave: D414080\\
	& M1 & Mixer& Marki Microwave: M1-0408\\
	& M2 & IQ mixer& Marki Microwave: IQ0307LXP\\
	& M3 & Mixer& Mini-Circuits: ZAD-8+\\
	& PS1 & Power splitter/combiner& Mini-Circuits: ZX10-2-71-s+\\
	& PS2 & Power splitter/combiner& Mini-Circuits: ZFRSC-2050+\\
	& S2 & Switch& Mini-Circuits: ZFSWA2-63DR+\\
\hline
\hline
\end{tabular}
\end{table*}

\end{document}